\documentclass[11pt,preprint]{aastex}
\slugcomment{{\sc Accepted to ApJ:} November 9, 2012} 

\usepackage{natbib}
\usepackage{rotating}
\usepackage{epsfig}
\usepackage{subfig}
\usepackage{color}
\newcommand{\arcs}{\mbox{\ensuremath{^{\prime\prime}}}} 

\shorttitle{The Stellar Populations and $M_{BH}$s of PSQs}
\shortauthors{Cales et al.}

\begin{document}

\title{The Properties of Post-Starburst Quasars Based on Optical Spectroscopy}



\author{Sabrina L. Cales\altaffilmark{1}, Michael S. Brotherton\altaffilmark{1}, Zhaohui Shang\altaffilmark{2,1}, Jessie C. Runnoe\altaffilmark{1},  Michael A. DiPompeo\altaffilmark{1}, Vardha Nicola Bennert\altaffilmark{3}, Gabriela Canalizo\altaffilmark{4,5}, Kyle D. Hiner\altaffilmark{4}, R. Stoll\altaffilmark{6}, Rajib Ganguly\altaffilmark{7}, Aleksandar Diamond-Stanic\altaffilmark{8}}


\altaffiltext{1}{Department of Physics and Astronomy, University of Wyoming, Laramie, WY 82071; scales@uwyo.edu, mbrother@uwyo.edu, shang@uwyo.edu, jrunnoe@uwyo.edu, mdipompe@uwyo.edu}
\altaffiltext{2}{Tianjin Normal University, Tianjin 300387, China, (+86) 22 23766527}
\altaffiltext{3}{Physics Department, California Polytechnic State University, San Luis Obispo, CA 93407; vbennert@calpoly.edu}
\altaffiltext{4}{Department of Physics and Astronomy, University of California, Riverside, CA 92521; gabriela.canalizo@ucr.edu, khine001@ucr.edu }
\altaffiltext{5}{Institute of Geophysics and Planetary Physics, University of California, Riverside, CA 92521} 
\altaffiltext{6}{Department of Astronomy, The Ohio State University, Columbus, OH 43210; stoll@astronomy.ohio-state.edu, (614) 292-1773}
\altaffiltext{7}{Department of Computer Science, Engineering, \& Physics, University of Michigan-Flint, Flint, MI 48502; ganguly@umflint.edu, (810)762-0787} 
\altaffiltext{8}{Center for Astrophysics and Space Sciences, University of California, San Diego; La Jolla, CA 92093; aleks@ucsd.edu; (858)534-2230}


\begin{abstract}

 We present optical spectroscopy of a sample of 38 post-starburst quasars (PSQs) at $z \sim 0.3$, 29 of which have morphological classifications based on \emph{Hubble Space Telescope} imaging. These broad-lined active galactic nuclei (AGNs) possess the spectral signatures of massive intermediate-aged stellar populations making them potentially useful for studying connections between nuclear activity and host galaxy evolution. We model the spectra in order to determine the ages and masses of the host stellar populations, and the black hole masses and Eddington fractions of the AGNs. Our model components include an instantaneous starburst, a power-law, and emission lines. We find the PSQs have $M_{BH} \sim\ 10^8\ M_{\odot}$ accreting at a few percent of Eddington luminosity and host $\sim10^{10.5}\ M_{\odot}$ stellar populations which are several hundred Myr to a few Gyr old. 
We investigate relationships among these derived properties, spectral properties, and morphologies. We find that PSQs hosted in spiral galaxies have significantly weaker AGN luminosities, older starburst ages, and narrow emission-line ratios diagnostic of ongoing star-formation when compared to their early-type counterparts. We conclude that the early-type PSQs are likely the result of major mergers and were likely luminous infrared galaxies in the past, while spiral PSQs with more complex star-formation histories are triggered by less dramatic events (e.g., harassment, bars). We provide diagnostics to distinguish the early-type and spiral hosts when high spatial resolution imaging is not available. 
  
\end{abstract}
\clearpage

\section{Introduction}
\label{sec:Intro}







  Galaxies harbor supermassive black holes (BHs) at their centers \citep[e.g.,][]{kormendyrichstone95}, the masses of which correlate with that of the host galaxies' bulge component \citep[$M_{BH} \sim 0.15\% M_{bulge}$;][]{merrittferrarese01, magorrian98, gebhardt00b}. An even stronger correlation exists between the BH mass and the bulge stellar velocity dispersion \citep{gebhardt00a, ferraresemerritt00, tremaine02}. These correlations suggests that BHs and their host galaxies have common evolutionary histories. The nature of the mechanism which is responsible for `coevolution' of both BH and bulge or even whether only one mechanism is responsible over cosmic time remains unclear. In order to grow to their observed masses, accreting black holes (i.e., AGN and/or quasars) are expected to be a phase in the life of every galaxy \citep{richstone98}.
    
    A new paradigm involving two mechanisms responsible for mutual BH-bulge growth has been suggested \citep[e.g.,][]{hasinger08,hopkinshernquist09,schawinski10a,bennert11}. In the early universe, major-merger driven evolution dominates and is responsible for producing the bulk of the brightest quasars at $z\ =\ 2-3$. Below $z\ \sim\ 1$, secular evolution and minor interactions become the main fueling mechanisms. Thus, at lower redshifts, less massive systems preferentially show activity \citep[i.e., AGN cosmic downsizing, e.g.,][]{heckman04}. Furthermore, transitioning from quasar to Seyfert luminosities, fueling rates and triggering mechanisms may also change \citep{hopkinshernquist09}, such that bars in spiral hosts may be sufficient to fuel the nuclear activity of Seyfert galaxies.
 
  In merger-driven evolutionary scenarios, mergers trigger starbursts and the ignition of AGN activity that can in turn inhibit both star formation and its own fueling through feedback (\citealt*{dimatteo05}; \citealt*{springel05}; \citealt{hopkins06}). A natural consequence of such models is the existence of objects that have luminous quasar activity, starburst or post-starburst signatures, along with indications of a recent merger, such as tidal debris. Indeed objects like this do exist, such as UN J1025$-$0040 which also been called a post-starburst quasar \citep[PSQ,][]{brotherton99, canalizo00, brotherton02}. 
  
The post-starburst classification is given for galaxies exhibiting strong Balmer absorption lines, indicating intense star formation in the past $\sim 1$ Gyr, and a lack of ongoing star formation, as indicated by having little or no nebular emission lines \citep{dresslergunn83}. Recently, the traditional definition for post-starburst galaxies has been found to be too narrow to encompass the full range in post-starburst populations \citep*{falkenberg09}. In particular, placing a limit on nebular emission lines (i.e., [\ion{O}{2}] $\lambda3727$) introduces a bias against AGNs \citep{yan06, wild09, kocevski11} and can cause gross underestimation of post-starburst galaxies hosting AGN since AGN can power significant emission from [\ion{O}{2}]. 

Several studies of AGN host galaxies in the Sloan Digital Sky Survey (SDSS) have shown that the most luminous AGN have had a burst of star formation in the past $\sim 1$ Gyr \citep{vandenberk01, kauffmann03}. 
Additionally, there exists mounting evidence that post-starburst signatures are enhanced in AGN compared to galaxies which do not exhibit AGN activity \citep{kocevski09, goto06, georgakakis08}. In particular, \citet{goto06} determines that the fraction of AGNs showing post-starburst features is at least 4.2\%, while the fraction of normal galaxies exhibiting these features is 0.2\%. 
  
  Low-redshift objects hosting the most massive starburst populations and most luminous AGN may be our best chance at finding the analogs of merger-induced systems at the quasar epoch. \citet[][hereafter C11]{cales11} studied the morphology and disturbance fraction of PSQs via \emph{HST}/ACS F606W imaging of the most luminous PSQ examples at $z\ \sim\ 0.3$ from a spectroscopically selected catalog \citep{brotherton07}. C11 find that PSQs are a heterogeneous population of early-type and spiral hosts, with disturbances being equally distributed among the morphologies. The presence of early-type hosts which appear to be major-merger remnants, along with spiral hosts, both isolated and with companions, is suggestive of a more complicated picture than can be explained by galaxy mergers alone. It has become evident that at least two mechanisms are responsible for triggering PSQs, consistent with  studies involving mutual BH-bulge growth. 
  
  Our follow-up project uses Keck and KPNO-4m optical spectroscopy of a sample of 38 post-starburst quasars (PSQs) at $z \sim 0.3$. We aim to characterize the BH masses, Eddington fractions, starburst masses, and starburst ages of PSQs via spectral modeling. Furthermore, 29 of these objects have morphological data from C11. We continue to characterize the fundamental properties of the AGN and starburst in PSQs and extend the study by investigating the interplay between the properties of PSQs and their morphological subpopulations. 

We describe our sample, selection, and data reductions in \S~\ref{sec:Data}. The methodology for decomposing AGN and starburst stellar populations along with the outputted results and derived fundamental AGN and post-starburst properties are given in \S~\ref{sec:Mod}. We present correlations between fitted and derived properties of the AGN and starburst features in addition to describing how the early-type and spiral host populations differ in \S~\ref{sec:Anal}. 
A summary of our results is given in \S~\ref{sec:Summ}. We adopt the cosmology $H_o = 70$ km s$^{-1}$Mpc$^{-1}$ and a flat universe where $\Omega_{M} = 0.3$ and $\Omega_{\Lambda} = 0.7$.

\section{Data}
\label{sec:Data}

\subsection{Sample}
\label{sec:Data.Catalog}

  We investigate a sample of 38 objects spectroscopically selected from the Sloan Digital Sky Survey data release 3 \citep[SDSS DR3;][]{abazajian05} that meet the following criteria: 
  
\begin{itemize}
\item Broad emission lines as defined by the SDSS DR3 online database (FWHM $> 1000$ km s$^{-1}$)
\item $r$ model magnitudes $\lesssim19$ 
\item 0.25 $< z < $0.45
\item $S/N > 8$ between rest-wavelengths of 4150 and 4250 \AA\ in the SDSS spectra
\item Summation of the Balmer absorption lines H$\delta$, H$\zeta$, and H$\eta$ $>$ 2 \AA\ rest-frame equivalent width at a significance greater than 6$\sigma$
\item H$\delta > 1$ \AA\  rest-frame equivalent width
\item Balmer Break $> 0.9$; based on the ratio of the fluxes at two 100 \AA\ wide regions centered at rest-wavelengths 4035 and 3790 \AA
\end{itemize}

The above criteria ensure that all objects in the sample are luminous AGN with clear post-starburst stellar populations. We call these objects PSQs regardless of whether their AGN component alone exceeds a formal luminosity separating quasar from Seyfert galaxy. Table~\ref{tab:sample} characterizes some observational properties of the sample. C11 classified morphologies and measured quasar-to-host light contributions based on \emph{HST}/ACS imaging of a subsample of 29 of these objects. The following sections provide details of our additional ground based spectroscopy which we obtained both in order to improve S/N and wavelength coverage compared to the SDSS spectra.

\begin{deluxetable}{clrrccc}
\tabletypesize{\scriptsize}
\tablecolumns{8}
\tablewidth{0pc}
\tablecaption{Journal of Observations \label{tab:sample}}
\tablehead{
\colhead{Object} & \colhead{$z$}    & \colhead{$r$\tablenotemark{a}} &  \colhead{$M_r$\tablenotemark{b}} & \colhead{Total Int.} & \colhead{Spectroscopy} & \colhead{Imaging\tablenotemark{c}}  \\
\colhead{SDSS}  & \colhead{} & \colhead{} & \colhead{} & \colhead{Time (s)}    & \colhead{} & \colhead{} }
\startdata
J003043.59$-$103517.6 & 0.296 & 18.26 & $-$22.98 & 1200  & Keck  & \emph{HST} \\
J005739.19$+$010044.9 & 0.253 & 17.61 & $-$23.23 & 1200  & Keck  & \emph{HST} \\
J015259.46$+$142738.0 & 0.311 & 18.44 & $-$22.87 & 1200  & Keck  & \ldots \\
J020258.94$-$002807.5 & 0.339 & 18.19 & $-$23.50 & 1200  & Keck  & \emph{HST} \\
J021447.00$-$003250.6 & 0.349 & 18.54 & $-$23.09 & 1200  & Keck  & \emph{HST} \\
J023253.42$-$082832.1 & 0.265 & 17.50 & $-$23.54 & 1200  & Keck  & \ldots \\
J023700.30$-$010130.5 & 0.344 & 18.58 & $-$23.05 & 1200  & Keck  & \emph{HST} \\
J025735.33$-$001631.3 & 0.362 & 18.67 & $-$22.95 & 1200  & Keck  & \ldots \\
J032143.15$-$064517.5 & 0.365 & 19.26 & $-$22.56 & 2400  & Keck  & \ldots \\
J040210.90$-$054630.3 & 0.270 & 18.70 & $-$22.45 & 1600  & Keck  & \emph{HST} \\
J074621.06$+$335040.7 & 0.284 & 17.97 & $-$23.12 & 1200  & Keck  & \emph{HST} \\
J075045.00$+$212546.3 & 0.408 & 18.01 & $-$24.18 & 1200  & Keck  & \emph{HST} \\
J075521.30$+$295039.2 & 0.334 & 18.69 & $-$22.83 & 1200  & Keck  & \emph{HST} \\
J075549.56$+$321704.1 & 0.420 & 18.82 & $-$23.24 & 2400  & Keck  & \emph{HST} \\
J081018.67$+$250921.2 & 0.263 & 17.41 & $-$23.20 & 300  & Keck  & \emph{HST} \\
J105816.81$+$102414.5 & 0.275 & 18.29 & $-$22.63 & 3600  & KPNO  & \emph{HST} \\
J115159.59$+$673604.8 & 0.274 & 18.44 & $-$22.64 & 7200  & KPNO  & \emph{HST} \\
J115355.58$+$582442.3 & 0.319 & 18.29 & $-$22.88 & 3600  & KPNO  & \emph{HST} \\
J123043.41$+$614821.8 & 0.324 & 18.63 & $-$22.77 & 7200  & KPNO  & \emph{HST} \\
J124833.52$+$563507.4 & 0.266 & 17.45 & $-$23.46 & 3600  & KPNO  & \emph{HST} \\
J140513.75$+$625008.2 & 0.386 & 18.59 & $-$24.15 & 3600  & KPNO  & \ldots \\
J145640.99$+$524727.2 & 0.277 & 18.14 & $-$22.84 & 3600  & KPNO  & \emph{HST} \\
J145658.15$+$593202.3 & 0.326 & 18.58 & $-$22.86 & 2400  & SDSS  & \emph{HST} \\
J154534.55$+$573625.1 & 0.268 & 18.07 & $-$22.88 & 3600  & KPNO  & \emph{HST} \\
J155214.85$+$565916.9 & 0.335 & 18.49 & $-$22.97 & 3600  & KPNO   & \ldots \\
J164444.92$+$423304.5 & 0.317 & 18.98 & $-$22.16 & 7200  & KPNO  & \emph{HST} \\
J170046.95$+$622056.4 & 0.276 & 18.58 & $-$22.51 & 3600  & KPNO  & \emph{HST} \\
J170819.80$+$603759.4 & 0.289 & 18.92 & $-$22.19 & 1200  & KPNO  & \ldots \\
J210200.42$+$000501.8 & 0.329 & 18.16 & $-$23.38 & 2400  & Keck  & \emph{HST} \\
J211343.20$-$075017.6 & 0.420 & 18.44 & $-$23.96 & 1200  & Keck  & \emph{HST} \\
J211838.12$+$005640.6 & 0.384 & 18.49 & $-$23.72 & 1200  & Keck  & \emph{HST} \\
J212843.42$+$002435.6 & 0.346 & 18.78 & $-$22.92 & 1200  & Keck  & \emph{HST} \\
J230614.18$-$010024.4 & 0.267 & 17.77 & $-$23.22 & 1200  & Keck  & \emph{HST} \\
J231055.50$-$090107.6 & 0.364 & 18.50 & $-$23.40 & 3600  & Keck  & \emph{HST} \\
J231317.85$-$082238.4 & 0.366 & 18.40 & $-$23.09 & 1200  & Keck  & \ldots \\
J233430.89$+$140649.7 & 0.363 & 18.57 & $-$23.26 & 1200  & Keck  & \emph{HST} \\
J234335.48$-$005758.1 & 0.341 & 18.00 & $-$23.74 & 1200  & Keck  & \ldots \\
J234403.55$+$154214.0 & 0.288 & 18.33 & $-$22.92 & 1200  & Keck  & \emph{HST} \\
\enddata
\tablenotetext{a}{SDSS DR7 $r$ AB magnitudes (modelMag\_r).}
\tablenotetext{b}{SDSS DR7 dereddened K-corrected $r$ absolute AB magnitudes. }
\tablenotetext{c}{Denotes whether morphological data is available from \emph{HST} ACS/WFC imaging.}
\end{deluxetable}


\subsection{KPNO Observations}
\label{sec:Data.KPNOSample}

  We obtained long-slit spectra of twelve PSQs and their companions using the Kitt Peak National Observatory (KPNO) 4 m Mayall telescope on the nights of 2006 May 20-23. We used the R-C spectrograph along with the 312 groove mm$^{-1}$ KPC-10A grating, in first order, producing a dispersion of 2.75 \AA\ pixel$^{-1}$ and a resolution of 6.9 \AA\ FWHM on the TK2B 2048 $\times$ 2048 CCD detector. The wavelength range covered is 3200 to 7200 \AA. The detector read noise and gain are 4 e$^{-}$ and 1.9 e$^{-}$ ADU$^{-1}$, respectively. The long slit is covered at a scale of 0\farcs69 pixel$^{-1}$; the slit width was 300 $\mu$m, or 2\arcs. 
  
  We applied a standard observation strategy with the exception of rotating the slit in order to observe the primary PSQ and its nearest bright companion within 20\arcs\ (to be presented in a future publication). We obtained three 1200 second exposures for each target PSQ. There were three instances where we observed the same target twice changing the slit angle in order to observe a different companion (i.e., SDSS~J115159.59$+$673604.8, SDSS~J123043.41$+$614821.8 and SDSS~J164444.92$+$423304.5). In these cases we obtained six 1200 second exposures of the primary PSQ. For one source (SDSS~J170819.80$+$603759.4), two of the three 1200-second exposures had low S/N due to poor conditions, and we use the data from a single exposure in our analysis.
  
  Reductions of the spectra were carried out in a standard manner using the IRAF\footnotemark\ software. After initial trimming, bias removal, and flat fielding, we extracted the one-dimensional cleaned spectrum using {\it apall}. We found the dispersion solution by identifying the lines of a He-Ne-Ar lamp. The spectra were flux calibrated using observations of several spectrophotometric standard stars \citep{massey88}. We median combined with S/N weights the individual spectrum using {\it scombine}. The combinations also include the SDSS spectrum in order to increase the S/N and the wavelength coverage of the resulting spectrum. Additional photometric corrections and combinations with other spectra were performed in a similar fashion as the Keck spectra and are described in Section~\ref{sec:Data.Final}. 
  
 \footnotetext{IRAF (Image Reduction and Analysis Facility) is distributed by the National Optical Astronomy Observatories, which are operated by AURA, Inc., under cooperative agreement with the National Science Foundation.}

\subsection{Keck Observations}
\label{sec:Data.KeckSample}

Spectroscopic observations were carried out on 2005 November 1 and 2 with the Low-Resolution Imaging Spectrometer \cite[LRIS;][]{oke95} on the Keck I telescope.  For the blue side (LRIS-B), we used the 600 groove mm$^{-1}$ grism blazed at 4000~\AA, yielding a dispersion of 0.63\,\AA\ pixel$^{-1}$.  For the red side (LRIS-R), we used the 400 groove mm$^{-1}$ grating blazed at 8500~\AA, yielding a dispersion of 1.86~\AA\ pixel$^{-1}$.  The slit was 1\arcsec\ wide, projecting to $\sim$7 pixels on the UV and blue-optimized CCD on LRIS-B and $\sim$5 pixels on the Tektronix 2048$\times$2048 CCD on LRIS-R.

We obtained between one or two exposures for each object, typically 1200~s each, dithering along the slit between exposures. The typical seeing for all observations was 0\farcs7 in $V$. Two or three spectrophotometric standards from \citet{massey88} were observed at parallactic angle each night for flux calibration.

The position angles (PAs) were selected so as to include nearby companion galaxies. The majority of objects were observed at low airmass to minimize the effects of differential atmospheric refraction.  The only object that was not observed near transit was SDSS~J040210.90$-$054630.3, which was observed at an airmass of 1.43. 

The spectra were reduced with IRAF, using standard reduction procedures.  After subtracting bias, dividing by a normalized halogen lamp flat-field frame and removing sky lines, we rectified the two-dimensional spectra and placed them on a wavelength scale using the least-mean-squares fit of cubic spline segments to identified lines in a Hg-Ne-Cd-Zn lamp. We calibrated the spectra using the spectrophotometric standards from \citet{massey88}.  The distortions in the spatial coordinate were removed with the IRAF {\it apextract} routines.  For each slit position, we had two or three individual frames;  we averaged the spatially corrected spectra using the IRAF task {\it scombine}.

\subsection{Finalized Spectra and Uncertainties}
\label{sec:Data.Final}

Since we chose to orient the slit to also observe neighboring objects we did not observe at parallactic angle. We used the SDSS spectra to correct for atmospheric diffraction slit losses. First, 
we divide our spectra by the corresponding SDSS spectra and fit the result by a low-order Legendre polynomial. We then corrected our spectra by dividing by the fitted polynomial. The resulting corrected spectra were well matched to the flux calibration of the SDSS spectra. 


We corrected for Galactic extinction using the \citet*{schlegel98} maps and the IRAF task \textit{deredden} which utilizes the \citet*{cardelliclaytonmathis89} extinction curves. We converted the spectra to rest-frame using the IRAF \textit{dopcor} task using SDSS redshifts. We note that we retained observed fluxes.


We have used a single uniform method to estimate the noise for each spectrum. For each pixel, we compute the root-mean-square value of the difference between the signal and the average using nine pixels centered on the input pixel. We use this estimated noise spectrum to weight individual points in the fitting procedure we describe below.

\section{Stellar Population Synthesis and AGN Component Modeling}
\label{sec:Mod}

   We utilized the IRAF task \textit{specfit} \citep{kriss94} to model the starburst populations and AGN contributions to the PSQs. The $\chi^2$ minimization technique of \textit{specfit} simultaneously models multiple components. We characterized the AGN contribution of the PSQs using a power-law, UV and optical iron emission line blends, as well as multiple Gaussian emission lines from the AGN broad and narrow-line regions. A \citet[][\textit{private communication}]{charlotbruzual07} instantaneous starburst (ISB) varying in age and mass describes the post-starburst component to the PSQs. We characterize the multiple fit components as either originating from the (1) AGN, (2) narrow-line regions, or (3) starburst, and we discuss these components in that order for the remainder of the paper.
   
\begin{figure}[tbhp]
\figurenum{1} 
\centering
\includegraphics[width=6in]{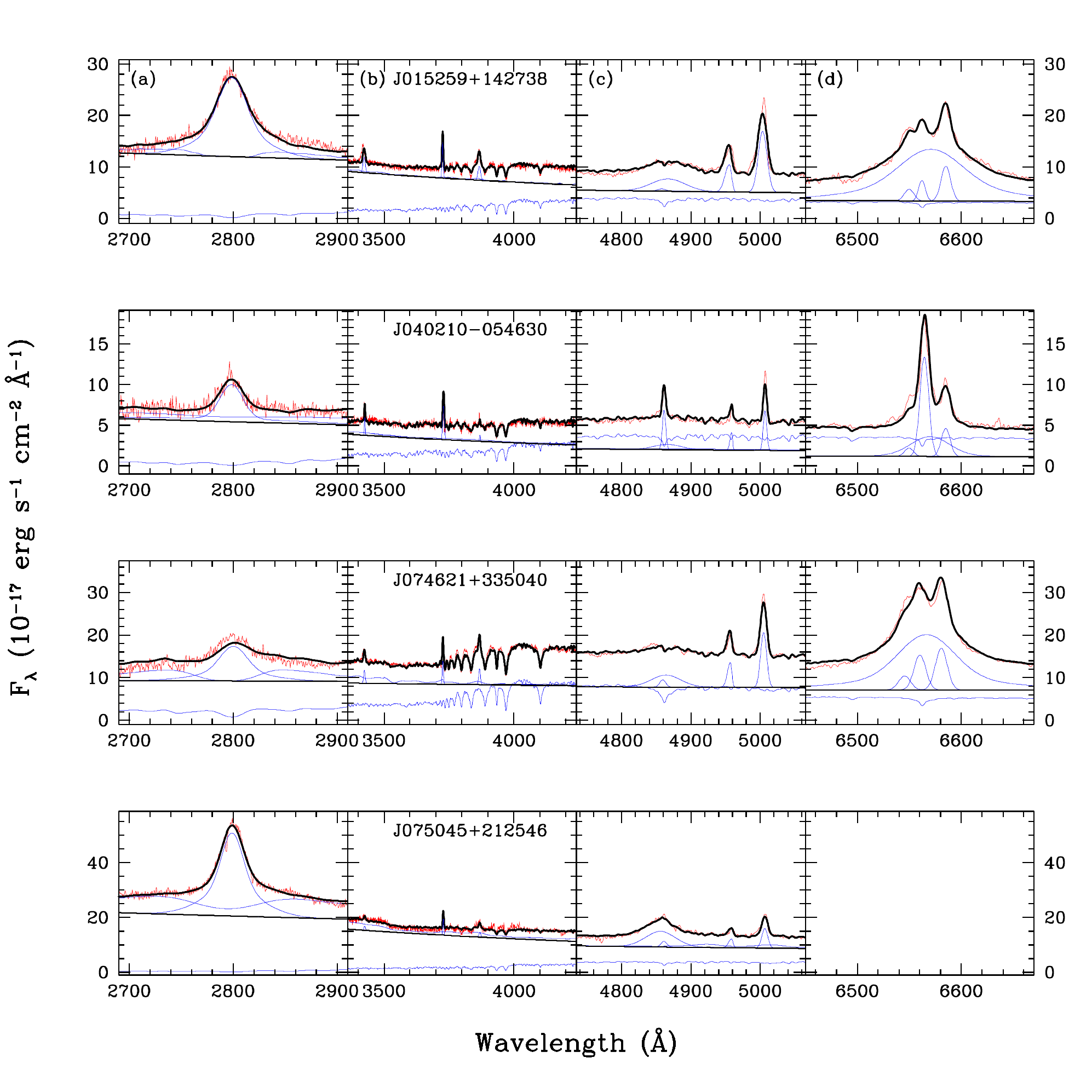}
\caption{Examples of our spectral decomposition of AGN and post-starburst stellar population highlighting the (a) Mg II, (b) Balmer break, (c) H$\beta$, and (d) H$\alpha$ regions. The red line is the data. The blue lines make up the components used in the fitting. For ease of interpretation, we plot the emission lines and Fe II templates above the power-law. The black line is the model fit to the data.\\ (A color version of this figure is available in the online journal.) \label{fig:zoom} }
\end{figure}

     We used the same initial guess parameters for each PSQ while changing the step sizes of the power-law index (0.02, 0.05, and 0.08) and starburst age (50, 100, and 200 Myr) corresponding to five runs for each object. We give plots of the runs corresponding to the best fits in the Appendix as well as zoom in on the broad lines and Balmer break for a few of our objects in Figure~\ref{fig:zoom}.
      
   Table~\ref{tab:AGN} gives the fitting results for the AGN power-law, \ion{Fe}{2} templates, and broad lines. The AGN power-law contribution is described by a normalization factor at 1000 \AA\ and the power-law index, $\alpha$: $f_{\lambda} \propto \lambda^{-\alpha}$. The UV and optical iron emission line blends are modeled using the UV and optical \ion{Fe}{2} templates derived from I Zw 1 \citep{vestergaardwilkes01, borosongreen92}. We convolved a Gaussian with the \ion{Fe}{2} templates to simulate different velocity widths. The \textit{specfit} task velocity-weight interpolates between the these templates and is also allowed to vary in intensity for each object.
   
   Gaussian emission lines are described by the flux under the profile, line centroid, full-width at half max (FWHM) and an asymmetry parameter. We found fitting two broad-line Gaussians to be satisfactory in matching the asymmetries and wings of emission lines originating from the broad-line region (i.e., \ion{Mg}{2}, H${\beta}$ and H${\alpha}$). Thus, we held the asymmetry parameter at `symmetric' and include two broad Gaussians in order to better model line asymmetries. We obtain a single broad-line FWHM by numerically finding the FWHM of the superposition of the two Gaussians. In order to separate the narrow and broad lines and also to characterize other narrow lines of interest we included a single Gaussian for each of the narrow lines; [\ion{Ne}{5}] $\lambda$3426, [\ion{O}{2}] $\lambda$3727, [\ion{Ne}{3}] $\lambda$3869, H${\beta}\ \lambda$4861, [\ion{O}{3}] $\lambda$4959, [\ion{O}{3}] $\lambda$5007, [\ion{N}{2]} $\lambda$6548, H${\alpha}\ \lambda$6563 and [\ion{N}{2}] $\lambda$6583. Table~\ref{tab:NLRa} and Table~\ref{tab:NLRb} give the fitting results for the narrow lines.
   
   The ISB modeling technique assumes that all stars are coeval with the same chemical composition. We used the models of \citet{charlotbruzual07} of solar metallicity and a \citet{chabrier03} initial mass function (IMF) to model the starburst contribution and find the best fit by varying the starburst age and intensity. The \textit{specfit} task age-weight interpolates between the stellar population ages of 56, 75, 100, 133, 177, 237, 316, 422, 562, 750, 1000, 1330, 1770, and 2370 Myr. Our age estimates are insensitive to changes in the IMF. We discuss starburst mass calculations in \S~\ref{sec:Mod.SBprop}. Table~\ref{tab:SB} gives fitting results for the post-starburst stellar populations.

   
    
   After choosing the fit corresponding to the best $\chi^2$, we analyzed the goodness-of-fit by eye for each object and interactively made adjustments to the fit parameters as necessary. We discarded components where the S/N was low ($\le 5$, particularly troublesome for \ion{Mg}{2} and in some objects the UV \ion{Fe}{2} templates) and in the complicated H$\beta$ region where degeneracies may exist between the emission of H${\beta}\ +\ $[\ion{O}{3}] $\lambda$4959$\ +\ $[\ion{O}{3}] $\lambda$5007 and the stellar absorption at H${\beta}$. We also note that for some objects the iron emission from I Zw 1 does not match the observed iron emission on the blue side of \ion{Mg}{2}. This mis-match is understood to be a result of different relative contributions to the multiplets giving rise to the iron emission. The fits to these objects were greatly improved if we simply doubled the flux of the template red-ward of 3025\AA.    
   
   We note that the \emph{HST} photometry of C11 matches the AGN power-law continuum to within 25\% in flux. This is in general good agreement considering variability issues and aperture differences.

\subsection{Caveats about Spectral Modeling}
\label{sec:Mod.MultSBprop}

  While our spectral fits are good in general, our model assumptions are too simple in some respects and their effects on our results should be considered more closely.  In particular, metallicity can vary between galaxies and within galaxies. Dust, often associated with starbursts and sometimes AGNs, can also be present along the line of sight and affect model parameters.  Finally, the original starburst event we model as an instantaneous burst certainly takes place in the environment of one or more galaxies with existing, older stellar populations. On-going star formation may be present as well. We briefly discuss these points below. However, each of these may represent more complex problems requiring more and higher S/N data to be more fully explored in future investigations.
 
  Metallicity affects the spectra of stars. If we assumed a metallicity higher than solar we would fit a stellar population that is younger and less massive. Conversely, assuming a lower metallicity would result in our fitting a stellar population that is older and more massive. In the case of the PSQ prototype UN J1025$-$0040, the uncertainty in the age of the 400 Myr old stellar population was $\pm 50$ Myrs assuming a reasonable range in metallicity \citep{brotherton99}. Their methodology is similar to ours and we expect similar uncertainties. 
  
 While our fits do not require reddening, dust may be present along the line of sight to the AGN and/or stellar component of our PSQs. Dust reddening makes a stellar population appear less luminous and older. Therefore, if significant dust is present the true stellar population age would be younger than our reported measurement and the mass larger. For a visual extinction of 0.1 magnitudes a 422 Myr population appears both older and less luminous by $\sim10$\%. The differences are less for older populations. The dust reddening laws toward AGN and starbursts typically differ further complicating more detailed modeling \citep{calzetti94, richards03}.   

 Our selection criteria ensure that an intermediate-age population is present and strong, but we also know that a more complex stellar population is likely present that includes both older and younger stars. We know that a few of the PSQs have morphologies that show knots of on-going star formation (C11), suggesting that young stellar populations may be present.  A similarly simple model, AGN$+$instantaneous burst, of the prototype PSQ, UN J1025-0040 \citep{brotherton99} required revising to include a younger stellar population when high-resolution blue Hubble Space Telescope images were obtained \citep{brotherton02}.  Because younger stellar populations have relatively weak spectral signatures and are difficult to identify with spectra alone in the presence of post-starburst populations, we want to suggest that this issue be kept in mind when interpreting spectral fitting results.

  The imaging also shows galaxies that appear to be spirals with bulges, which typically have older populations, and ellipticals that appear to be post-merger remnants.  The dominant intermediate age populations are most likely created as part of the merging process, but the older remain present.  Older populations do have spectral signatures that are distinct (e.g. \ion{Mg}{1} b absorption, larger \ion{Ca}{2} H \& K ratios relative to Balmer line absorption), so we have a better chance of identifying objects that require a more complex stellar population in our fitting. \ion{Mg}{1} b absorption is not readily apparent in individual spectra, but we note that it is visible in composite PSQ spectra we have examined.

  We performed an additional set of fits in which we also included an older stellar population with an age of 5.6 Gyrs.  This additional component did not result in improved fits (as evaluated by $\chi^2$ and by eye examining the absorption lines in detail).  This could be the result of degeneracies inherent in fitting many components to a complex spectrum, something worth additional study, or because the older components do not contribute significantly to the observed flux.  This does not mean that they are not present, only that the fits may be degenerate or that the older components are weak.  Others have investigated the issues involving fitting multiple-aged stellar components to a spectrum with a single-aged stellar synthesis model \citep[e.g.,][]{serratrager07,tragersomerville09}, who have characterized some of the resulting biases and limitations.

  We quantitatively investigated this effect ourselves with our own techniques by simulating spectra of a  post-starburst plus old stellar component for a range of flux/mass ratios.  We created our simulated spectra by combining ratios of 422 Myr old and 5.6 Gyr old Charlot \& Bruzual instantaneous burst models and adding artificial noise consistent with our data.  We then fit the resulting simulated spectra with a single-aged stellar component.  The intermediate age stellar age and mass were recovered to better than 10\% as long as the older population was less than 70\% of the stellar mass.  At higher mass fractions, the intermediate age population fit was compromised and skewed the results to indicate older, more massive populations.   The AGN component additionally greatly complicates the task and can mask the presence of different-aged stellar components.  We conclude that our results are robust as long as the post-starburst population is massive enough to be dominant in flux, as it is at least in some cases \citep[e.g.,][]{hiner12}, but that higher S/N spectra (letting us measure the \ion{Mg}{1} b feature more reliably) or another waveband of high-resolution imaging \citep[e.g.,][]{brotherton02} will be required to definitively resolve this issue.
   
 We additionally note that in our simulations, when a fit overestimates a post-starburst stellar population age, the equivalent widths of the Balmer lines are greatly overestimated (at least $\sim 30$\%), although as a complication the AGN component can dilute the equivalent widths. In our single intermediate-age population fits, the Balmer lines of the data are well characterized by our model and for only one object (SDSS J023700.30$-$010130.5) we note that there might be a significant overestimation.  The good fits overall and the fact that we've used a single, consistent approach to fitting all objects suggests that even in the presence of some systematic issues, relative measurements are still likely meaningful and correlation analysis of interest.
 
 The issue of how to robustly fit complex spectra involving multiple stellar components of unknown metallicity, an AGN, and unknown and perhaps complex dust reddening, at a range of S/N ratios, is complicated and deserving of additional deeper investigation and is beyond the scope of the present work.  Our existing data are well and consistently fit by our simple model, and to to investigate these more sophisticated models would require additional data. Even with more data, degeneracies may make it extremely difficult to find unique and perfect solutions.  Again, our fits appear good and likely provide useful information about the PSQs, but keep in mind these caveats and how they may bias the results when interpreting our measurements.

\begin{deluxetable}{lcccccccccccccccccc}
\tabletypesize{\tiny}
\tablecolumns{19}
\tablewidth{0pc}
\tablecaption{AGN Fitting Results \label{tab:AGN}}
\tablehead{
\colhead{Object} & \multicolumn{2}{c}{Power Law} & \colhead{ } & \multicolumn{2}{c}{UV Fe} &  \colhead{ } & \multicolumn{2}{c}{Fe II} &   \colhead{ } & \multicolumn{2}{c}{Mg II} &   \colhead{ } & \multicolumn{2}{c}{H$\beta$}  & \colhead{ } & \multicolumn{2}{c}{H$\alpha$}  & \colhead{H$\alpha \rightarrow$H$\beta$\tablenotemark{f}} \\
\cline{2-3} \cline{5-6} \cline{8-9} \cline{11-12} \cline{14-15} \cline{17-18} 
\colhead{SDSSJ}  & \colhead{Norm\tablenotemark{a}} & \colhead{$\alpha$\tablenotemark{b}} & \colhead{ } & \colhead{Scale\tablenotemark{c}} & \colhead{FWHM\tablenotemark{d}} & \colhead{ } & \colhead{Scale} & \colhead{FWHM} & \colhead{ } & \colhead{Flux\tablenotemark{e}}  & \colhead{FWHM} & \colhead{ } & \colhead{Flux} & \colhead{FWHM} & \colhead{ } & \colhead{Flux} & \colhead{FWHM} & \colhead{FWHM} }
\startdata
J003043$-$103517 &       8.19 &       0.35 & &    \ldots\tablenotemark{g} &         \ldots & &    \ldots &         \ldots & &    \ldots &         \ldots & &     106.48 &         7617 & &     782.66 &         7926 &         9639 \\
J005739$+$010044 &       8.09 &       0.28 & &    \ldots &         \ldots & &    \ldots &         \ldots & &     175.26 &         7665 & &    \ldots &         \ldots & &     497.10 &         4310 &         5084 \\
J015259$+$142738 &      55.51 &       1.49 & &       0.17 &         1792 & &    \ldots &         \ldots & &     739.39 &         4128 & &     148.29 &         3552 & &    1025.44 &         3747 &         4389 \\
J020258$-$002807 &       4.60 &       0.07 & &       0.37 &         9746 & &    \ldots &         \ldots & &    \ldots &         \ldots & &    \ldots &         \ldots & &    \ldots &         \ldots &         \ldots \\
J021447$-$003250 &      17.77 &       0.86 & &       0.14 &         3669 & &       0.20 &         4770 & &     266.95 &         3779 & &      88.32 &         5193 & &    \ldots &         \ldots &         \ldots \\
J023253$-$082832 &      16.30 &       0.38 & &       0.04 &        11881 & &    \ldots &         \ldots & &     478.03 &         5078 & &     385.69 &         7077 & &    1863.20 &         7596 &         9218 \\
J023700$-$010130 &      40.81 &       1.96 & &       0.48 &        10058 & &    \ldots &         \ldots & &     409.88 &         6846 & &    \ldots &         \ldots & &    \ldots &         \ldots &         \ldots \\
J025735$-$001631 &      16.64 &       0.83 & &       0.09 &         1855 & &       0.36 &         2128 & &     268.56 &         2132 & &     113.85 &         1872 & &    \ldots &         \ldots &         \ldots \\
J032143$-$064517 &      49.66 &       1.95 & &       0.15 &         5515 & &    \ldots &         \ldots & &     228.08 &         4469 & &     120.45 &         5694 & &    \ldots &         \ldots &         \ldots \\
J040210$-$054630 &      34.85 &       1.81 & &       0.18 &        11729 & &    \ldots &         \ldots & &     192.84 &         2943 & &      35.12 &         2897 & &     131.00 &         2037 &         2315 \\
J074621$+$335040 &      16.73 &       0.46 & &       0.06 &         5391 & &    \ldots &         \ldots & &     374.56 &         3845 & &     153.24 &         3079 & &    1174.74 &         3244 &         3773 \\
J075045$+$212546 &      91.06 &       1.45 & &       1.32 &         7961 & &       0.69 &         2147 & &    1220.57 &         3300 & &     310.65 &         3139 & &    \ldots &         \ldots &         \ldots \\
J075521$+$295039 &       6.06 &       0.48 & &       0.24 &         9406 & &    \ldots &         \ldots & &     188.69 &         2328 & &    \ldots &         \ldots & &    \ldots &         \ldots &         \ldots \\
J075549$+$321704 &      43.29 &       1.49 & &       0.38 &         8578 & &       0.21 &         4715 & &     267.67 &         3794 & &     172.21 &         4505 & &    \ldots &         \ldots &         \ldots \\
J081018$+$250921 &     114.36 &       1.06 & &       4.24 &         9618 & &       2.05 &         9477 & &    6808.27 &         8397 & &    1100.02 &         5259 & &    1682.93 &         4357 &         5142 \\
J105816$+$102414 &      47.41 &       1.40 & &    \ldots &         \ldots & &    \ldots &         \ldots & &     660.14 &         8426 & &     310.05 &         9099 & &     857.88 &         4683 &         5548 \\
J115159$+$673604 &       7.25 &       0.82 & &    \ldots &         \ldots & &    \ldots &         \ldots & &    \ldots &         \ldots & &    \ldots &         \ldots & &     845.27 &         4781 &         5669 \\
J115355$+$582442 &      24.73 &       1.19 & &       0.54 &         3850 & &       0.52 &         4799 & &     544.21 &         2837 & &     125.73 &         2710 & &     541.56 &         3466 &         4044 \\
J123043$+$614821 &      28.25 &       1.16 & &       0.64 &         8865 & &       0.28 &         4687 & &     753.57 &         3464 & &      84.21 &         3623 & &     990.86 &         4226 &         4980 \\
J124833$+$563507 &       7.90 &       0.22 & &       0.39 &         5999 & &       0.76 &         4237 & &    \ldots &         \ldots & &     193.10 &         4645 & &    1351.19 &         5125 &         6098 \\
J140513$+$625008 &       3.84 &       0.23 & &       0.23 &         8434 & &       0.22 &         5496 & &     543.69 &         4336 & &    \ldots &         \ldots & &     558.75 &         4505 &         5325 \\
J145640$+$524727 &      11.55 &       0.75 & &       0.36 &        10286 & &       0.22 &         4331 & &    \ldots &         \ldots & &      79.52 &         4084 & &     353.92 &         3266 &         3800 \\
J145658$+$593202 &       5.03 &       0.53 & &    \ldots &         \ldots & &    \ldots &         \ldots & &    \ldots &         \ldots & &      55.34 &         4686 & &     453.90 &         6826 &         8240 \\
J154534$+$573625 &      45.24 &       1.13 & &       0.47 &        11684 & &    \ldots &         \ldots & &    1257.66 &         8640 & &     288.11 &        10263 & &    2144.13 &         8093 &         9852 \\
J155214$+$565916 &      10.71 &       0.71 & &       0.20 &        10251 & &    \ldots &         \ldots & &    \ldots &         \ldots & &      61.05 &         3830 & &     334.88 &         5032 &         5982 \\
J164444$+$423304 &      11.05 &       0.75 & &       0.56 &        11129 & &       0.27 &         6715 & &     200.53 &         2503 & &      58.49 &         2512 & &     319.85 &         2482 &         2848 \\
J170046$+$622056 &      38.50 &       1.67 & &    \ldots &         \ldots & &    \ldots &         \ldots & &    1098.66 &         9929 & &    \ldots &         \ldots & &     900.31 &         6724 &         8109 \\
J170819$+$603759 &       8.56 &       0.64 & &    \ldots &         \ldots & &    \ldots &         \ldots & &    \ldots &         \ldots & &      89.78 &         5173 & &    \ldots &         \ldots &         \ldots \\
J210200$+$000501 &     136.70 &       2.35 & &       0.45 &        11819 & &    \ldots &         \ldots & &     883.42 &         4471 & &     138.74 &         4330 & &    \ldots &         \ldots &         \ldots \\
J211343$-$075017 &     101.42 &       2.16 & &    \ldots &         \ldots & &    \ldots &         \ldots & &    1525.69 &         5977 & &     185.33 &         7055 & &    \ldots &         \ldots &         \ldots \\
J211838$+$005640 &      46.35 &       1.58 & &       0.32 &         4175 & &    \ldots &         \ldots & &     645.84 &         4539 & &     201.13 &         7200 & &    \ldots &         \ldots &         \ldots \\
J212843$+$002435 &      13.22 &       0.49 & &       0.00 &         3671 & &    \ldots &         \ldots & &     363.04 &         2983 & &     165.69 &         3203 & &    \ldots &         \ldots &         \ldots \\
J230614$-$010024 &      10.43 &       0.25 & &    \ldots &         \ldots & &    \ldots &         \ldots & &     412.81 &         4728 & &     156.91 &         4440 & &     651.67 &         2688 &         3097 \\
J231055$-$090107 &      50.07 &       2.10 & &       0.43 &         9152 & &    \ldots &         \ldots & &     805.11 &         5628 & &    \ldots &         \ldots & &    \ldots &         \ldots &         \ldots \\
J231317$-$082238 &      11.75 &       0.57 & &       0.25 &         3190 & &       0.25 &         4478 & &     430.44 &         4872 & &     137.26 &         4322 & &    \ldots &         \ldots &         \ldots \\
J233430$+$140649 &      15.78 &       0.51 & &       0.34 &        10045 & &    \ldots &         \ldots & &     588.16 &         6233 & &     174.42 &         6635 & &    \ldots &         \ldots &         \ldots \\
J234335$-$005758 &       3.46 &       0.07 & &    \ldots &         \ldots & &    \ldots &         \ldots & &    \ldots &         \ldots & &      45.45 &         7095 & &    \ldots &         \ldots &         \ldots \\
J234403$+$154214 &      31.76 &       1.98 & &       0.55 &         9162 & &    \ldots &         \ldots & &     423.99 &         4664 & &       6.14 &        20023 & &     746.51 &         4393 &         5187 \\
\enddata
\tablenotetext{a}{The flux at 1000 \AA, in units of 10$^{-17}$ ergs s$^{-1}$ cm$^{-2}$ \AA$^{-1}$.}
\tablenotetext{b}{Power-law index.}
\tablenotetext{c}{The factor by which the Fe II templates have been scaled.}
\tablenotetext{d}{FWHM are given in rest-frame km s$^{-1}$.}
\tablenotetext{e}{Scaling for the Gaussian emission lines corresponds to the integrated area under the curve as flux in units of 10$^{-17}$ ergs s$^{-1}$ cm$^{-2}$.}
\tablenotetext{f}{Estimated values of H$\beta$ based on H$\alpha$ measurements using the relation given by \citet{shen11}.}
\tablenotetext{g}{Denotes discarded measurements or no coverage. }
\end{deluxetable}

\begin{deluxetable}{lccccccccccccccccc}
\tabletypesize{\tiny}
\tablecolumns{18}
\tablewidth{0pc}
\tablecaption{Narrow-Line Fitting Results - A\label{tab:NLRa}}
\tablehead{
\colhead{Object} & \multicolumn{2}{c}{[Ne V]} & \colhead{ } & \multicolumn{2}{c}{[O II]} &  \colhead{ } & \multicolumn{2}{c}{[Ne III]} &   \colhead{ } & \multicolumn{2}{c}{H$\beta$} &   \colhead{ } & \multicolumn{2}{c}{[O III] $\lambda$4959}  & \colhead{ } & \multicolumn{2}{c}{[O III] $\lambda$5007}  \\
\cline{2-3} \cline{5-6} \cline{8-9} \cline{11-12} \cline{14-15} \cline{17-18} 
\colhead{SDSSJ}  & \colhead{Flux\tablenotemark{a}} & \colhead{FWHM\tablenotemark{b}} & \colhead{ } & \colhead{Flux} & \colhead{FWHM} & \colhead{ } & \colhead{Flux} & \colhead{FWHM} & \colhead{ } & \colhead{Flux}  & \colhead{FWHM} & \colhead{ } & \colhead{Flux} & \colhead{FWHM} & \colhead{ } & \colhead{Flux} & \colhead{FWHM}}
\startdata
J003043$-$103517 &       9.29 &          798 & &      51.62 &          731 & &      21.77 &          798 & &      30.58 &          798 & &      22.19 &          798 & &      66.56 &          798 \\
J005739$+$010044 &       6.95 &          469 & &      24.64 &          411 & &      11.47 &          469 & &      43.85 &          469 & &      12.46 &          469 & &      37.37 &          469 \\
J015259$+$142738 &      33.24 &          898 & &      41.37 &          449 & &      32.15 &          898 & &       6.78 &          898 & &      63.05 &          898 & &     189.16 &          898 \\
J020258$-$002807 &       7.93 &          590 & &     152.68 &          562 & &      -1.13 &          590 & &      78.70 &          590 & &      20.24 &          590 & &      60.71 &          590 \\
J021447$-$003250 &      13.65 &          752 & &      28.59 &          440 & &      15.42 &          752 & &      27.79 &          752 & &      28.98 &          752 & &      86.95 &          752 \\
J023253$-$082832 &      15.32 &          699 & &      36.25 &          754 & &      21.02 &          699 & &      11.25 &          699 & &      18.73 &          699 & &      56.18 &          699 \\
J023700$-$010130 &      13.42 &          503 & &      81.27 &          523 & &       3.20 &          503 & &      53.77 &          503 & &      31.65 &          503 & &      94.94 &          503 \\
J025735$-$001631 &      19.29 &          517 & &      28.07 &          458 & &      11.16 &          517 & &      16.15 &          517 & &      32.66 &          517 & &      97.98 &          517 \\
J032143$-$064517 &       9.26 &          621 & &       7.73 &          792 & &      16.88 &          621 & &    \ldots &         \ldots & &      14.29 &          621 & &      42.86 &          621 \\
J040210$-$054630 &       3.55 &          348 & &      32.00 &          586 & &       9.23 &          348 & &      29.15 &          348 & &      10.16 &          348 & &      30.49 &          348 \\
J074621$+$335040 &      31.47 &          612 & &      32.46 &          449 & &      22.90 &          612 & &      17.67 &          612 & &      46.58 &          612 & &     139.75 &          612 \\
J075045$+$212546 &      15.45 &          608 & &      30.39 &          389 & &      11.46 &          608 & &      21.34 &          608 & &      25.65 &          608 & &      76.96 &          608 \\
J075521$+$295039 &       7.69 &          398 & &      55.86 &          426 & &       4.64 &          398 & &      41.49 &          398 & &      22.17 &          398 & &      66.52 &          398 \\
J075549$+$321704 &      16.08 &          695 & &      19.61 &          736 & &      10.84 &          695 & &      18.09 &          695 & &      24.20 &          695 & &      72.60 &          695 \\
J081018$+$250921 &      23.92 &          532 & &    \ldots &         \ldots & &      36.57 &          532 & &    \ldots &         \ldots & &      65.71 &          532 & &     197.13 &          532 \\
J105816$+$102414 &      13.79 &          429 & &      38.29 &          943 & &      15.22 &          429 & &      17.97 &          429 & &      42.88 &          429 & &     128.63 &          429 \\
J115159$+$673604 &       9.49 &          584 & &      23.87 &          507 & &       6.38 &          584 & &      13.60 &          584 & &      15.66 &          584 & &      46.97 &          584 \\
J115355$+$582442 &      18.44 &          659 & &      58.94 &          572 & &       5.89 &          659 & &      19.63 &          659 & &      36.92 &          659 & &     110.75 &          659 \\
J123043$+$614821 &       8.49 &          449 & &      41.79 &          553 & &      10.41 &          449 & &      23.88 &          449 & &      12.83 &          449 & &      38.48 &          449 \\
J124833$+$563507 &    \ldots &         \ldots & &      17.68 &          440 & &    \ldots &         \ldots & &    \ldots &         \ldots & &    \ldots &         \ldots & &    \ldots &         \ldots \\
J140513$+$625008 &      22.15 &          798 & &      30.04 &          617 & &      14.63 &          798 & &      11.82 &          798 & &      24.66 &          798 & &      73.97 &          798 \\
J145640$+$524727 &    \ldots &         \ldots & &      23.00 &          490 & &    \ldots &         \ldots & &       6.38 &          200 & &    \ldots &         \ldots & &    \ldots &         \ldots \\
J145658$+$593202 &      11.02 &          384 & &      50.49 &          516 & &       7.15 &          384 & &      22.66 &          384 & &      21.31 &          384 & &      63.92 &          384 \\
J154534$+$573625 &      43.09 &          561 & &     157.47 &          728 & &      28.11 &          561 & &      47.83 &          561 & &     100.47 &          561 & &     301.40 &          561 \\
J155214$+$565916 &       4.42 &          328 & &      34.27 &          606 & &       3.84 &          328 & &      18.71 &          328 & &      10.31 &          328 & &      30.94 &          328 \\
J164444$+$423304 &    \ldots &         \ldots & &      28.66 &          779 & &       1.55 &          691 & &      16.94 &          691 & &      10.82 &          691 & &      32.45 &          691 \\
J170046$+$622056 &      23.89 &          753 & &      41.60 &         1144 & &      16.56 &          753 & &      34.35 &          753 & &      18.11 &          753 & &      54.33 &          753 \\
J170819$+$603759 &    \ldots &         \ldots & &    \ldots &         \ldots & &    \ldots &         \ldots & &      28.04 &          716 & &      26.50 &          716 & &      79.51 &          716 \\
J210200$+$000501 &      23.88 &          814 & &      50.77 &          631 & &       2.93 &          814 & &       4.95 &          814 & &      11.97 &          814 & &      35.91 &          814 \\
J211343$-$075017 &      38.08 &          688 & &      55.43 &          538 & &      25.91 &          688 & &    \ldots &         \ldots & &      47.50 &          688 & &     142.51 &          688 \\
J211838$+$005640 &      19.99 &          671 & &      22.14 &          450 & &      11.42 &          671 & &       9.02 &          671 & &      46.84 &          671 & &     140.51 &          671 \\
J212843$+$002435 &      11.07 &          642 & &      14.59 &          693 & &      10.13 &          642 & &       5.04 &          642 & &      16.70 &          642 & &      50.10 &          642 \\
J230614$-$010024 &      11.91 &          344 & &      61.32 &          473 & &       6.99 &          344 & &      69.20 &          344 & &      21.36 &          344 & &      64.08 &          344 \\
J231055$-$090107 &      17.28 &          674 & &      56.33 &          627 & &      10.46 &          674 & &      25.68 &          674 & &      20.59 &          674 & &      61.77 &          674 \\
J231317$-$082238 &       9.32 &          794 & &       9.00 &          300 & &       5.34 &          794 & &       8.13 &          794 & &      17.92 &          794 & &      53.76 &          794 \\
J233430$+$140649 &      30.54 &          689 & &     140.41 &          525 & &      15.50 &          689 & &      55.65 &          689 & &      64.98 &          689 & &     194.94 &          689 \\
J234335$-$005758 &       8.37 &          603 & &      42.66 &          818 & &       6.16 &          603 & &      35.25 &          603 & &      16.57 &          603 & &      49.71 &          603 \\
J234403$+$154214 &    \ldots &         \ldots & &      10.98 &          729 & &    \ldots &         \ldots & &    \ldots &         \ldots & &    \ldots &         \ldots & &    \ldots &         \ldots \\
\enddata
\tablenotetext{a}{Scaling for the Gaussian emission lines corresponds to the integrated area under the curve as flux in units of 10$^{-17}$ ergs s$^{-1}$ cm$^{-2}$.}
\tablenotetext{b}{FWHM are given in rest-frame km s$^{-1}$.}
\tablecomments{The FWHM of the narrow emission lines should not be considered physical due to the low spectral resolution of our KPNO and Keck spectra.}
\end{deluxetable}

\begin{deluxetable}{lcccccccc}
\tabletypesize{\tiny}
\tablecolumns{9}
\tablewidth{0pc}
\tablecaption{Narrow-Line Fitting Results - B\label{tab:NLRb}}
\tablehead{
\colhead{Object} &\multicolumn{2}{c}{[N II] $\lambda$6548}  & \colhead{ } & \multicolumn{2}{c}{H$\alpha$}  & \colhead{ } & \multicolumn{2}{c}{[N II] $\lambda$6583}  \\
\cline{2-3} \cline{5-6} \cline{8-9} 
\colhead{SDSSJ}  & \colhead{Flux\tablenotemark{a}} & \colhead{FWHM\tablenotemark{b}} & \colhead{ } & \colhead{Flux} & \colhead{FWHM} & \colhead{ } & \colhead{Flux} & \colhead{FWHM} }
\startdata
J003043$-$103517 &      37.67 &          694 & &     122.86 &          694 & &     113.01 &          694 \\
J005739$+$010044 &      33.70 &          469 & &     224.43 &          469 & &     101.10 &          469 \\
J015259$+$142738 &      26.10 &          493 & &      32.62 &          354 & &      78.29 &          493 \\
J020258$-$002807 &    \ldots &         \ldots & &    \ldots &         \ldots & &    \ldots &         \ldots \\
J021447$-$003250 &    \ldots &         \ldots & &    \ldots &         \ldots & &    \ldots &         \ldots \\
J023253$-$082832 &      21.08 &          699 & &     102.62 &          699 & &      63.24 &          699 \\
J023700$-$010130 &    \ldots &         \ldots & &    \ldots &         \ldots & &    \ldots &         \ldots \\
J025735$-$001631 &    \ldots &         \ldots & &    \ldots &         \ldots & &    \ldots &         \ldots \\
J032143$-$064517 &    \ldots &         \ldots & &    \ldots &         \ldots & &    \ldots &         \ldots \\
J040210$-$054630 &      11.00 &          456 & &     130.00 &          456 & &      37.00 &          456 \\
J074621$+$335040 &      46.50 &          612 & &     117.05 &          612 & &     139.49 &          612 \\
J075045$+$212546 &    \ldots &         \ldots & &    \ldots &         \ldots & &    \ldots &         \ldots \\
J075521$+$295039 &    \ldots &         \ldots & &    \ldots &         \ldots & &    \ldots &         \ldots \\
J075549$+$321704 &    \ldots &         \ldots & &    \ldots &         \ldots & &    \ldots &         \ldots \\
J081018$+$250921 &      17.00 &          532 & &      11.33 &          532 & &      51.01 &          532 \\
J105816$+$102414 &      20.33 &          429 & &      49.17 &          429 & &      60.98 &          429 \\
J115159$+$673604 &      12.48 &          295 & &      41.61 &          295 & &      37.44 &          295 \\
J115355$+$582442 &      35.49 &          659 & &      88.45 &          659 & &     106.48 &          659 \\
J123043$+$614821 &      41.68 &          449 & &     217.61 &          449 & &     125.04 &          449 \\
J124833$+$563507 &      25.94 &          513 & &      83.91 &          513 & &      77.81 &          513 \\
J140513$+$625008 &      18.78 &          798 & &      37.76 &          798 & &      56.34 &          798 \\
J145640$+$524727 &       9.42 &          200 & &      44.01 &          200 & &      28.26 &          200 \\
J145658$+$593202 &      17.87 &          384 & &      97.65 &          384 & &      53.62 &          384 \\
J154534$+$573625 &      92.60 &          561 & &     211.54 &          561 & &     277.80 &          561 \\
J155214$+$565916 &      16.43 &          328 & &      63.16 &          328 & &      49.28 &          328 \\
J164444$+$423304 &      11.53 &          691 & &      93.17 &          691 & &      34.59 &          691 \\
J170046$+$622056 &      11.11 &          455 & &      49.99 &          455 & &      33.32 &          455 \\
J170819$+$603759 &    \ldots &         \ldots & &    \ldots &         \ldots & &    \ldots &         \ldots \\
J210200$+$000501 &    \ldots &         \ldots & &    \ldots &         \ldots & &    \ldots &         \ldots \\
J211343$-$075017 &    \ldots &         \ldots & &    \ldots &         \ldots & &    \ldots &         \ldots \\
J211838$+$005640 &    \ldots &         \ldots & &    \ldots &         \ldots & &    \ldots &         \ldots \\
J212843$+$002435 &    \ldots &         \ldots & &    \ldots &         \ldots & &    \ldots &         \ldots \\
J230614$-$010024 &      31.65 &          344 & &     202.85 &          344 & &      94.96 &          344 \\
J231055$-$090107 &    \ldots &         \ldots & &    \ldots &         \ldots & &    \ldots &         \ldots \\
J231317$-$082238 &    \ldots &         \ldots & &    \ldots &         \ldots & &    \ldots &         \ldots \\
J233430$+$140649 &    \ldots &         \ldots & &    \ldots &         \ldots & &    \ldots &         \ldots \\
J234335$-$005758 &    \ldots &         \ldots & &    \ldots &         \ldots & &    \ldots &         \ldots \\
J234403$+$154214 &       9.56 &          406 & &      42.66 &          406 & &      28.67 &          406 \\
\enddata
\tablenotetext{a}{Scaling for the Gaussian emission lines corresponds to the integrated area under the curve as flux in units of 10$^{-17}$ ergs s$^{-1}$ cm$^{-2}$.}
\tablenotetext{b}{FWHM are given in rest-frame km s$^{-1}$.}
\tablecomments{The FWHM of the narrow emission lines should not be considered physical due to the low spectral resolution of our KPNO and Keck spectra.}
\end{deluxetable}

\begin{deluxetable}{lrrrrr}
\tabletypesize{\tiny}
\tablecolumns{6}
\tablewidth{0pc}
\tablecaption{Starburst Properties \label{tab:SB}}
\tablehead{
\colhead{Object} &  \colhead{Age}  & \colhead{Raw\tablenotemark{a}}  & \colhead{log Mass}   & \colhead{log $L_{SB}$\tablenotemark{b}}   & \colhead{log $L_{Tot}$\tablenotemark{c}} \\
\colhead{SDSS}  & \colhead{(Myr)}  & \colhead{Scale}  & \colhead{($M_{\sun}$)}  & \colhead{(erg s$^{-1}$)}   & \colhead{(erg s$^{-1}$)} }
\startdata
J003043$-$103517 &     740 &      26.28 &      10.27 &      43.47 &      43.85 \\
J005739$+$010044 &    1180 &      34.19 &      10.41 &      43.41 &      43.76 \\
J015259$+$142738 &    1080 &      22.78 &      10.42 &      43.44 &      43.94 \\
J020258$-$002807 &     970 &      32.25 &      10.62 &      43.67 &      43.98 \\
J021447$-$003250 &    1090 &      22.17 &      10.54 &      43.54 &      43.98 \\
J023253$-$082832 &    1690 &      18.80 &      10.36 &      43.15 &      43.81 \\
J023700$-$010130 &    1230 &      36.26 &      10.79 &      43.74 &      43.92 \\
J025735$-$001631 &     280 &      30.05 &      10.13 &      43.70 &      44.06 \\
J032143$-$064517 &    2330 &      22.37 &      10.93 &      43.52 &      43.87 \\
J040210$-$054630 &    2400 &      25.11 &      10.66 &      43.26 &      43.55 \\
J074621$+$335040 &     620 &      48.50 &      10.41 &      43.68 &      44.04 \\
J075045$+$212546 &    2330 &      25.51 &      11.11 &      43.69 &      44.39 \\
J075521$+$295039 &     790 &      23.89 &      10.39 &      43.54 &      43.83 \\
J075549$+$321704 &    2310 &      21.06 &      11.06 &      43.64 &      44.14 \\
J081018$+$250921 &      90 &      62.48 &       9.58 &      43.58 &      44.28 \\
J105816$+$102414 &    1180 &      51.56 &      10.68 &      43.67 &      43.96 \\
J115159$+$673604 &    1490 &      31.75 &      10.57 &      43.44 &      43.63 \\
J115355$+$582442 &     760 &      38.21 &      10.52 &      43.71 &      43.97 \\
J123043$+$614821 &    1340 &      21.92 &      10.54 &      43.46 &      43.92 \\
J124833$+$563507 &    1400 &      67.37 &      10.83 &      43.75 &      43.97 \\
J140513$+$625008 &     830 &      34.36 &      10.73 &      43.85 &      44.05 \\
J145640$+$524727 &    1890 &      21.89 &      10.52 &      43.25 &      43.66 \\
J145658$+$593202 &     960 &      17.13 &      10.30 &      43.36 &      43.67 \\
J154534$+$573625 &    2380 &      44.53 &      10.89 &      43.51 &      43.95 \\
J155214$+$565916 &    1210 &      16.23 &      10.41 &      43.37 &      43.81 \\
J164444$+$423304 &     470 &      23.92 &      10.11 &      43.49 &      43.83 \\
J170046$+$622056 &    2670 &      28.28 &      10.78 &      43.34 &      43.66 \\
J170819$+$603759 &    1270 &      20.73 &      10.37 &      43.32 &      43.68 \\
J210200$+$000501 &     790 &      54.88 &      10.73 &      43.89 &      44.11 \\
J211343$-$075017 &     750 &      52.35 &      10.97 &      44.12 &      44.34 \\
J211838$+$005640 &    1160 &      41.20 &      10.95 &      43.91 &      44.18 \\
J212843$+$002435 &    1690 &      18.88 &      10.65 &      43.42 &      43.88 \\
J230614$-$010024 &     890 &      48.31 &      10.50 &      43.62 &      43.95 \\
J231055$-$090107 &    2230 &      31.48 &      11.05 &      43.67 &      43.90 \\
J231317$-$082238 &    1500 &      28.73 &      10.85 &      43.69 &      44.07 \\
J233430$+$140649 &     240 &      18.19 &       9.84 &      43.49 &      44.12 \\
J234335$-$005758 &    1400 &      36.17 &      10.84 &      43.73 &      43.95 \\
J234403$+$154214 &    2570 &      34.88 &      10.90 &      43.47 &      43.64 \\
\enddata
\tablenotetext{a}{Scale directly from fitting.}
\tablenotetext{b}{Total integrated light of the starburst component from 3000 \AA\ to 6000 \AA.}
\tablenotetext{c}{Total integrated light (AGN plus starburst components) from 3000 \AA\ to 6000 \AA.}
\end{deluxetable}

\subsection{AGN Properties} 
\label{sec:Mod.AGNprop}

  From the fitting results, we were able to estimate two fundamental physical AGN properties: the black hole mass and Eddington ratio. We employed scaling relations that extend the reverberation mapping results of physical properties (i.e., $M_{BH}$ and radius of the broad-line region) to the observable quantities (i.e., continuum luminosity and broad line FWHM) of single-epoch data to calculate black hole mass. We used \ion{Mg}{2}, H${\beta}$, H${\alpha}$ as well as estimated values of H${\beta}$ based on the measurements of H${\alpha}$, being careful to discard data of low quality. All of our spectra have coverage at 5100 \AA. We used the 5100 \AA\ monochromatic luminosities based on the power-law components used in the fits as our AGN luminosities for the scaling relations below.
 
   From the relation given by \citet{vestergaardosmer09}, we estimated $M_{BH}$ based on the \ion{Mg}{2} broad line FWHM and the 5100 \AA\ monochromatic luminosity:
\begin{equation}
\label{eqn:vo09}
\frac{M_{BH}\ (\textrm{\ion{Mg}{2}})}{{M}_{\odot}} = 10^{6.96}\left[\frac{\textrm{FWHM \ion{Mg}{2}}}{1000\ \textrm{km s}^{-1}}\right]^2\left[\frac{\lambda L_{\lambda}(5100\ \textrm{\AA})}{10^{44}\ \textrm{ergs s}^{-1}}\right]^{0.5}.
\end{equation}
The scatter in the zero point of the relation is given to be 0.55 dex.

   We used the relation:
\begin{equation}
\label{eqn:vp06}
\frac{M_{BH}\ (\textrm{H}{\beta})}{M_{\odot}} = 10^{6.91\pm0.02} \left[ \frac{\textrm{FWHM H}{\beta}}{1000\ \textrm{km s}^{-1}} \right]^2 \left[ \frac{\lambda L_{\lambda}(5100\ \textrm{\AA})}{10^{44}\ \textrm{ergs s}^{-1}} \right]^{0.5}
\end{equation}
to calculate the black hole mass based on the H${\beta}$ broad line FWHM and the 5100 \AA\ AGN monochromatic luminosity \citep{vestergaardpeterson06}. The intrinsic scatter in the sample is described by 0.43 dex. We have given the relation in linear space as opposed to log space for ease of reading. We note that the relations given for \ion{Mg}{2} and H${\beta}$ are on the same mass scale.

   We estimated the black hole masses using H${\alpha}$ broad line FWHM and the 5100 \AA\ AGN monochromatic luminosity according to the \citet*{greene10} scaling relation: 
\begin{equation}
\label{eqn:g10}
\frac{M_{BH}\ (\textrm{H}{\alpha})}{{M}_{\odot}} = (9.7 \pm 0.5) \times 10^6 \left[\frac{\textrm{FWHM H}{\alpha}}{1000\ \textrm{km s}^{-1}}\right]^{2.06 \pm 0.06} \left[\frac{\lambda L_{\lambda}(5100\ \textrm{\AA})}{10^{44}\ \textrm{ergs s}^{-1}}\right]^{0.519 \pm 0.07}.
\end{equation}
Though the scatter in this relation is not reported, typical values are $\sim0.4-0.5$ dex. 

   Since stellar absorption from the post-starburst population contaminates broad emission in H${\beta}$, making the fits less reliable, and this stellar contamination is negligible for H$\alpha$ we used H${\alpha}$ as a proxy for  H${\beta}$ when H${\alpha}$ is present in the spectrum \citep{shen11}:
\begin{equation}
\label{eqn:s11}
\log \left( \frac{\textrm{FWHM H}{\beta}}{\textrm{km s}^{-1}} \right) = (-0.11 \pm 0.03) + (1.05 \pm 0.01) \times \log \left( \frac{\textrm{FWHM H}{\alpha}}{\textrm{km s}^{-1}} \right).
\end{equation}
We then used the estimated H${\beta}$ values in conjunction with Equation~\ref{eqn:vp06} to calculate black hole measurements for H${\beta}$ based on H${\alpha}$. There are slight differences between the two formalisms of \citet{greene10} and \citet{vestergaardpeterson06}; most notably between the prefactors and radius-luminosity (scaling of $\lambda L_{\lambda}$) relations. However, the differences are small; for our sample the difference is less than 8\%, much less than the intrinsic scatter.



   We have given several measurements of the black hole mass.  Each line has its own particular issue. The blue side of the spectra suffer from S/N degradation which sometimes gives unreliable measurements for \ion{Mg}{2}. Occasionally, up to $\sim50$\% of the H$\beta$ broad emission can be contaminated by stellar absorption. Due to the redshift range of our sample, sometimes H$\alpha$ is out of our observing window. In future sections we make comparisons using these values individually. However, we also give an adopted $M_{BH}$ by applying the following prescription. When we have three reliable measurements we adopt the median of these values. For two good measurements, we give the mean of the values. If there is only one reliable value, we adopt this as our $M_{BH}$. There is one object (SDSS J020258.94$-$002807.5) for which $H\alpha$ was not covered and both \ion{Mg}{2} and H$\beta$ were unreliable. 

   We estimated the AGN bolometric luminosity using the measured rest-frame flux at 5100 \AA\ from the fit to the power-law continuum and convert it into a total emitted luminosity using luminosity distances for our cosmology and an appropriate bolometric correction \citep*[$f$ = 8.1;][]{runnoe12}:
\begin{equation}
\label{eqn:Lbol}
L_{bol} = 4\pi D_L^2 f (1+z) \lambda F_{\lambda}(5100\textrm{\AA}).
\end{equation}
The Eddington ratio is given by $L_{bol}/L_{Edd}$ where we use $L_{Edd}$ $=$ 1.51 $\times$ 10$^{38}$($M_{BH}$/M$_{\odot}$) ergs s$^{-1}$ \citep{krolik99}. Table~\ref{tab:AGNcalc} gives black hole masses for each broad line measurement and their Eddington ratios.

\begin{figure}[tbhp]
\figurenum{2} 
\centering
\includegraphics[scale=0.6]{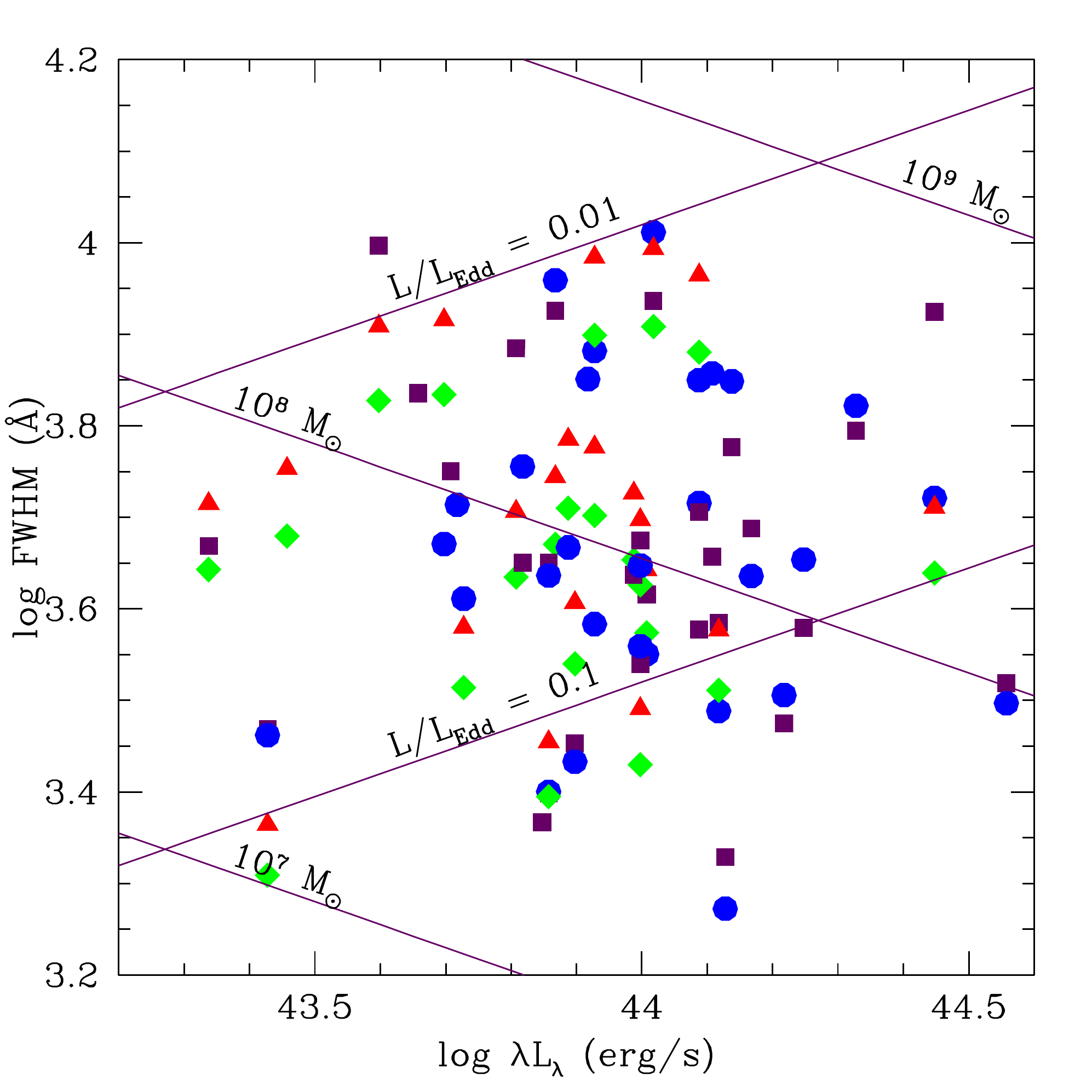}
\caption{FWHM versus $\lambda L_{\lambda}$(5100 \AA) based on measurements from Mg II (purple squares), H$\beta$ (blue circles), H$\alpha$ (green diamonds), and H$\beta$ predicted from H$\alpha$ (red triangles). Lines of constant $M_{BH}$ and $L/L_{Edd}$ are shown for Mg II. PSQs have black hole masses of $\sim10^8$ M$_{\odot}$  and Eddington fractions of several percent. \\ (A color version of this figure is available in the online journal.) \label{fig:FWHMvL} }
\end{figure}

  Figure~\ref{fig:FWHMvL} shows where the different broad line width measurements (i.e., \ion{Mg}{2}, H$\beta$, H$\alpha$ and predicted H$\beta$) lie as a function of  $\lambda L_{\lambda}$(5100 \AA). For reference we indicate lines of constant $M_{BH}$ and $L_{bol}/L_{Edd}$ based on \ion{Mg}{2}. Constant lines of $M_{BH}$ and $L_{bol}/L_{Edd}$ based on  H$\beta$ and H$\alpha$ will vary slightly in slope and intercept.

\begin{deluxetable}{lccccccccccccc}
\tabletypesize{\tiny}
\tablecolumns{14}
\tablewidth{0pc}
\tablecaption{AGN Properties \label{tab:AGNcalc}}
\tablehead{
\colhead{Object} & \multicolumn{5}{c}{log $M_{BH}$ ($M_{\sun}$)} & \colhead{ } & \multicolumn{5} {c}{$L/L_{Edd}$} &   \colhead{$f_{nuc}$\tablenotemark{a}} &  \colhead{log $L_{AGN}$\tablenotemark{b}} \\
\cline{2-6} \cline{8-12} 
\colhead{SDSS}  & \colhead{Mg II}  & \colhead{H$\beta$}  & \colhead{H$\alpha$} & \colhead{H$\alpha \rightarrow$H$\beta$}  & \colhead{Adopted}   & \colhead{ }  & \colhead{Mg II}   & \colhead{H$\beta$}  & \colhead{H$\alpha$} & \colhead{H$\alpha \rightarrow$H$\beta$} & \colhead{Adopted} &  \colhead{ } &  \colhead{(erg s$^{-1}$)} }
\startdata
J003043$-$103517 &    \ldots &       8.64 &       8.80 &       8.84 &       8.73 & &   \ldots &      0.011 &      0.007 &      0.007 &      0.009 &      0.581 &      43.61 \\
J005739$+$010044 &       8.36 &    \ldots &       8.19 &       8.23 &       8.29 & &      0.015 &   \ldots &      0.022 &      0.020 &      0.018 &      0.552 &      43.50 \\
J015259$+$142738 &       7.93 &       8.02 &       8.17 &       8.20 &       8.02 & &      0.065 &      0.053 &      0.037 &      0.035 &      0.053 &      0.687 &      43.78 \\
J020258$-$002807 &    \ldots &    \ldots &    \ldots &    \ldots &       \ldots & &   \ldots &   \ldots &   \ldots &   \ldots & \ldots &      0.502 &      43.68 \\
J021447$-$003250 &       7.89 &       8.39 &    \ldots &    \ldots &       8.21 & &      0.085 &      0.027 &   \ldots &   \ldots &      0.041 &      0.640 &      43.79 \\
J023253$-$082832 &       8.15 &       8.65 &       8.85 &       8.88 &       8.65 & &      0.047 &      0.015 &      0.009 &      0.009 &      0.015 &      0.778 &      43.70 \\
J023700$-$010130 &       8.19 &    \ldots &    \ldots &    \ldots &       8.19 & &      0.016 &   \ldots &   \ldots &   \ldots &      0.016 &      0.342 &      43.46 \\
J025735$-$001631 &       7.41 &       7.52 &    \ldots &    \ldots &       7.47 & &      0.278 &      0.218 &   \ldots &   \ldots &      0.244 &      0.567 &      43.82 \\
J032143$-$064517 &       7.90 &       8.33 &    \ldots &    \ldots &       8.17 & &      0.044 &      0.016 &   \ldots &   \ldots &      0.024 &      0.552 &      43.61 \\
J040210$-$054630 &       7.34 &       7.55 &       7.33 &       7.35 &       7.34 & &      0.065 &      0.041 &      0.068 &      0.064 &      0.065 &      0.484 &      43.24 \\
J074621$+$335040 &       7.92 &       7.95 &       8.10 &       8.12 &       7.95 & &      0.085 &      0.080 &      0.056 &      0.053 &      0.080 &      0.563 &      43.79 \\
J075045$+$212546 &       8.01 &       8.18 &    \ldots &    \ldots &       8.11 & &      0.192 &      0.128 &   \ldots &   \ldots &      0.153 &      0.800 &      44.29 \\
J075521$+$295039 &       7.35 &    \ldots &    \ldots &    \ldots &       7.35 & &      0.169 &   \ldots &   \ldots &   \ldots &      0.169 &      0.487 &      43.52 \\
J075549$+$321704 &       7.97 &       8.34 &    \ldots &    \ldots &       8.19 & &      0.101 &      0.043 &   \ldots &   \ldots &      0.060 &      0.685 &      43.97 \\
J081018$+$250921 &       8.76 &       8.57 &       8.54 &       8.56 &       8.57 & &      0.026 &      0.040 &      0.044 &      0.042 &      0.040 &      0.801 &      44.19 \\
J105816$+$102414 &       8.48 &       8.76 &       8.30 &       8.33 &       8.48 & &      0.013 &      0.007 &      0.020 &      0.018 &      0.013 &      0.482 &      43.64 \\
J115159$+$673604 &    \ldots &    \ldots &       8.11 &       8.15 &       8.11 & &   \ldots &   \ldots &      0.012 &      0.011 &      0.012 &      0.359 &      43.19 \\
J115355$+$582442 &       7.55 &       7.73 &       8.05 &       8.07 &       7.73 & &      0.121 &      0.080 &      0.038 &      0.036 &      0.080 &      0.460 &      43.63 \\
J123043$+$614821 &       7.77 &       8.03 &       8.27 &       8.30 &       8.03 & &      0.091 &      0.050 &      0.028 &      0.027 &      0.050 &      0.649 &      43.73 \\
J124833$+$563507 &    \ldots &       8.19 &       8.39 &       8.43 &       8.30 & &   \ldots &      0.027 &      0.017 &      0.016 &      0.021 &      0.403 &      43.58 \\
J140513$+$625008 &       7.96 &    \ldots &       8.33 &       8.36 &       8.18 & &      0.057 &   \ldots &      0.025 &      0.023 &      0.034 &      0.377 &      43.63 \\
J145640$+$524727 &    \ldots &       8.00 &       7.90 &       7.93 &       7.95 & &   \ldots &      0.029 &      0.036 &      0.033 &      0.032 &      0.609 &      43.44 \\
J145658$+$593202 &    \ldots &       8.10 &       8.55 &       8.59 &       8.38 & &   \ldots &      0.021 &      0.008 &      0.007 &      0.011 &      0.513 &      43.38 \\
J154534$+$573625 &       8.57 &       8.94 &       8.86 &       8.90 &       8.86 & &      0.015 &      0.006 &      0.008 &      0.007 &      0.008 &      0.644 &      43.76 \\
J155214$+$565916 &    \ldots &       8.04 &       8.40 &       8.43 &       8.25 & &   \ldots &      0.042 &      0.018 &      0.017 &      0.025 &      0.643 &      43.62 \\
J164444$+$423304 &       7.41 &       7.64 &       7.72 &       7.75 &       7.64 & &      0.148 &      0.088 &      0.072 &      0.069 &      0.088 &      0.541 &      43.56 \\
J170046$+$622056 &       8.48 &    \ldots &       8.48 &       8.53 &       8.48 & &      0.007 &   \ldots &      0.007 &      0.006 &      0.007 &      0.528 &      43.39 \\
J170819$+$603759 &    \ldots &       8.20 &    \ldots &    \ldots &       8.20 & &   \ldots &      0.018 &   \ldots &   \ldots &      0.018 &      0.556 &      43.42 \\
J210200$+$000501 &       7.92 &       8.11 &    \ldots &    \ldots &       8.03 & &      0.046 &      0.030 &   \ldots &   \ldots &      0.036 &      0.391 &      43.70 \\
J211343$-$075017 &       8.31 &       8.68 &    \ldots &    \ldots &       8.53 & &      0.036 &      0.016 &   \ldots &   \ldots &      0.022 &      0.394 &      43.94 \\
J211838$+$005640 &       8.06 &       8.68 &    \ldots &    \ldots &       8.47 & &      0.060 &      0.014 &   \ldots &   \ldots &      0.023 &      0.469 &      43.85 \\
J212843$+$002435 &       7.75 &       8.03 &    \ldots &    \ldots &       7.91 & &      0.158 &      0.082 &   \ldots &   \ldots &      0.108 &      0.652 &      43.69 \\
J230614$-$010024 &       8.04 &       8.20 &       7.87 &       7.89 &       8.04 & &      0.049 &      0.033 &      0.072 &      0.068 &      0.049 &      0.535 &      43.68 \\
J231055$-$090107 &       8.05 &    \ldots &    \ldots &    \ldots &       8.05 & &      0.025 &   \ldots &   \ldots &   \ldots &      0.025 &      0.413 &      43.51 \\
J231317$-$082238 &       8.15 &       8.27 &    \ldots &    \ldots &       8.21 & &      0.056 &      0.043 &   \ldots &   \ldots &      0.049 &      0.591 &      43.84 \\
J233430$+$140649 &       8.45 &       8.72 &    \ldots &    \ldots &       8.60 & &      0.041 &      0.022 &   \ldots &   \ldots &      0.029 &      0.764 &      44.00 \\
J234335$-$005758 &    \ldots &       8.57 &    \ldots &    \ldots &       8.57 & &   \ldots &      0.012 &   \ldots &   \ldots &      0.012 &      0.406 &      43.56 \\
J234403$+$154214 &       7.70 &       9.18 &       7.97 &       8.01 &       7.97 & &      0.023 &      0.001 &      0.013 &      0.011 &      0.013 &      0.326 &      43.15 \\
\enddata
\tablenotetext{a}{Integrated ratio of AGN to total light from 3000 \AA\ to 6000 \AA.}
\tablenotetext{b}{Total integrated light of the AGN power-law from 3000 \AA\ to 6000 \AA.}
\end{deluxetable}

\subsection{Post-Starburst Properties} 
\label{sec:Mod.SBprop}

   The post-starburst stellar populations can be described using two fundamental physical parameters, starburst age and mass. The post-starburst ages are known directly from the fitting results. The scale factor is a function of the mass and age of the starburst. The inputted PSQ spectra are in units of 10$^{-17}$ erg s$^{-1}$ cm$^{2}$ \AA$^{-1}$ while the template spectra are in units of L$_{\sun}$/\AA\ and scaled up by their age in Myr. Thus, we can derive the total starburst mass by using the scale factor and the outputted age: 
   
\begin{equation}
\label{eqn:Msb}
\frac{M_{SB}}{{M}_{\odot}}=  \textrm{Raw scale} \times \textrm{Age (Myr)} \times 10^{-17} \textrm{ erg s$^{-1}$ cm$^{2}$ \AA$^{-1}$} \times \left[\frac { 3.826 \times 10^{33} \textrm{ erg s$^{-1}$ \AA$^{-1}$} }{4\pi D_L^2 (1+z)}\right]^{-1}.
\end{equation}
The expression in brackets is the solar flux at the luminosity distance, $D_L$. Thus, the ratio between the scaled model and the solar flux at $D_L$ gives the starburst mass. Table~\ref{tab:SB} gives the starburst fitting results, derived starburst masses, and integrated luminosities.

\section{Results and Discussion}
\label{sec:Anal}

   We seek to explore joint AGN and starburst activity by investigating the interplay between the fundamental properties of PSQs and their morphological subpopulations. In \S~\ref{sec:Anal.Corr} we investigate the relationships between the fundamental properties of PSQs. We investigate the fundamental properties of PSQs in relation to their morphological subpopulations in \S~\ref{sec:Anal.Pop}. Finally, in \S~\ref{sec:Anal.Mode}, we give a simple prescription to reliably classify the morphology of PSQs based on spectral properties.

\begin{deluxetable}{llrrc}
\tabletypesize{\scriptsize}
\tablecolumns{5}
\tablewidth{0pc}
\tablecaption{Significant Correlations \label{tab:corr}}
\tablehead{
\colhead{Property 1} & \colhead{Property 2} & \colhead{$\rho$} & \colhead{P (\%)} & \colhead{Number} }
\startdata
\hline 
 \multicolumn{5}{c}{Total Population} \\
 \hline 
$z$ & $L_{Tot}$ &         0.518 &         0.086 &           38 \\
$z$ & $L_{SB}$ &         0.494 &         0.161 &           38 \\
$z$ & SB Mass &         0.428 &         0.737 &           38 \\
$\lambda L_{\lambda}$(5100 \AA) & $L/L_{Edd}$ Adopted &         0.433 &         0.740 &           37 \\
$\lambda L_{\lambda}$(5100 \AA) & [O III]/H$\beta$ &         0.481 &         0.527 &           32 \\
$\lambda L_{\lambda}$(5100 \AA) & [O III]/[O II] &         0.476 &         0.508 &           33 \\
$\alpha$ & SB Mass &         0.479 &         0.237 &           38 \\
$L_{Tot}$ & [O III]/H$\beta$ &         0.509 &         0.292 &           32 \\
$L_{Tot}$ & [N II]/H$\alpha$ &         0.678 &         0.073 &           21 \\
$L_{AGN}$ & $L_{SB}$ &         0.441 &         0.553 &           38 \\
$L_{AGN}$ & $L/L_{Edd}$ Adopted &         0.424 &         0.890 &           37 \\
$L_{AGN}$ & [O III]/H$\beta$ &         0.563 &         0.079 &           32 \\
$L_{AGN}$ & [O III]/[O II] &         0.493 &         0.359 &           33 \\
$L_{SB}$ & SB Mass &         0.452 &         0.438 &           38 \\
$L_{SB}$ & [N II]/H$\alpha$ &         0.549 &         0.990 &           21 \\
$L_{SB}$ & [Ne V]/[Ne III] &         0.504 &         0.276 &           33 \\
$M_{BH}$ Mg II & $L/L_{Edd}$ Mg II &        -0.742 &         0.000 &           29 \\
$M_{BH}$ H$\beta$ & $L/L_{Edd}$ H$\beta$ &        -0.809 &         0.000 &           30 \\
$M_{BH}$ H$\alpha$ & $L/L_{Edd}$ H$\alpha$ &        -0.739 &         0.013 &           21 \\
$M_{BH}$ H$\alpha$-H$\beta$ & $L/L_{Edd}$ H$\alpha$-H$\beta$ &        -0.739 &         0.013 &           21 \\
$M_{BH}$ Adopted & $L/L_{Edd}$ Adopted &        -0.724 &         0.000 &           37 \\
SB Mass & SB Age &         0.631 &         0.002 &           38 \\
$$[O III]/H$\beta$ & [N II]/H$\alpha$ &         0.838 &         0.003 &           17 \\
$$[O III]/H$\beta$ & [O III]/[O II] &         0.737 &         0.000 &           31 \\
$$[N II]/H$\alpha$ & [O III]/[O II] &         0.757 &         0.043 &           17 \\
\hline 
 \multicolumn{5}{c}{Elliptical Population} \\ 
 \hline 
$M_{BH}$ Mg II & $L/L_{Edd}$ Mg II &        -0.790 &         0.222 &           12 \\
$M_{BH}$ H$\beta$ & $L/L_{Edd}$ H$\beta$ &        -0.841 &         0.032 &           13 \\
$M_{BH}$ Adopted & $L/L_{Edd}$ Adopted &        -0.692 &         0.873 &           13 \\
SB Mass & SB Age &         0.747 &         0.333 &           13 \\
\hline 
 \multicolumn{5}{c}{Spiral Population} \\ 
 \hline 
$\alpha$ & SB Mass &         0.824 &         0.053 &           13 \\
$\alpha$ & SB Age &         0.764 &         0.238 &           13 \\
$M_{BH}$ Adopted & $L/L_{Edd}$ Adopted &        -0.753 &         0.298 &           13 \\
SB Mass & SB Age &         0.813 &         0.072 &           13 \\
\hline 
 \multicolumn{5}{c}{Undisturbed Population} \\ 
 \hline 
$\alpha$ & SB Mass &         0.824 &         0.005 &           17 \\
$L_{Tot}$ & [O III]/[O II] &         0.650 &         0.871 &           15 \\
$L_{AGN}$ & $L_{SB}$ &         0.674 &         0.301 &           17 \\
$L_{AGN}$ & [O III]/[O II] &         0.657 &         0.777 &           15 \\
$M_{BH}$ Mg II & $L/L_{Edd}$ Mg II &        -0.829 &         0.024 &           14 \\
$M_{BH}$ H$\beta$ & $L/L_{Edd}$ H$\beta$ &        -0.934 &         0.000 &           14 \\
$M_{BH}$ H$\alpha$ & $L/L_{Edd}$ H$\alpha$ &        -0.800 &         0.311 &           11 \\
$M_{BH}$ H$\alpha$-H$\beta$ & $L/L_{Edd}$ H$\alpha$-H$\beta$ &        -0.800 &         0.311 &           11 \\
$M_{BH}$ Adopted & $L/L_{Edd}$ Adopted &        -0.814 &         0.007 &           17 \\
SB Mass & SB Age &         0.620 &         0.792 &           17 \\
SB Mass & [O III]/[O II] &         0.646 &         0.921 &           15 \\
$$[O III]/H$\beta$ & [O III]/[O II] &         0.842 &         0.016 &           14 \\
\enddata
\tablenotetext{a}{$\lambda L_{\lambda}$(5100 \AA) is the monochromatic AGN luminosity at 5100 \AA\ in erg s$^{-1}$.}
\end{deluxetable}
\clearpage

\subsection{Correlations Based on Fundamental PSQ Properties} 
\label{sec:Anal.Corr}

   We calculate Spearman-rank correlation coefficient matrices involving several hundred correlation tests between fitted and derived parameters for (1) the total sample, (2) the early-type and spiral morphological classifications, and, (3) the disturbed and undisturbed classifications. We compute the probability of the correlations arising by chance and list those less than 1\% in Table~\ref{tab:corr}.
   
   
   We find a number of strong correlations among quasar and starburst properties, however many of these are due to selection effects which we will discuss below. Some merger induced evolutionary scenarios predict correlations among these properties of a more physical origin \citep{dimatteo05}. We find very few additional correlations of this type. 
   
   
   
   

   Our selection criteria provides an efficient means for selecting the brightest systems with both AGN and post-starburst populations. However, this then creates some artificial correlations between parameters. For example, the starburst and AGN must be comparable in luminosity in order for one to not swamp out light from the other, thus giving rise to a significant  correlation of the luminosities of both. We note that the mean AGN-to-total light, $f_{nuc}$, is 0.55 with only a standard deviation of 0.13. This leaves us with a limited parameter space to explore.
   


\subsubsection{Derived AGN Properties} 
\label{sec:Anal.CorrAGN}

\begin{figure}[tbhp]
\figurenum{3} 
\centering
\includegraphics[scale=0.6]{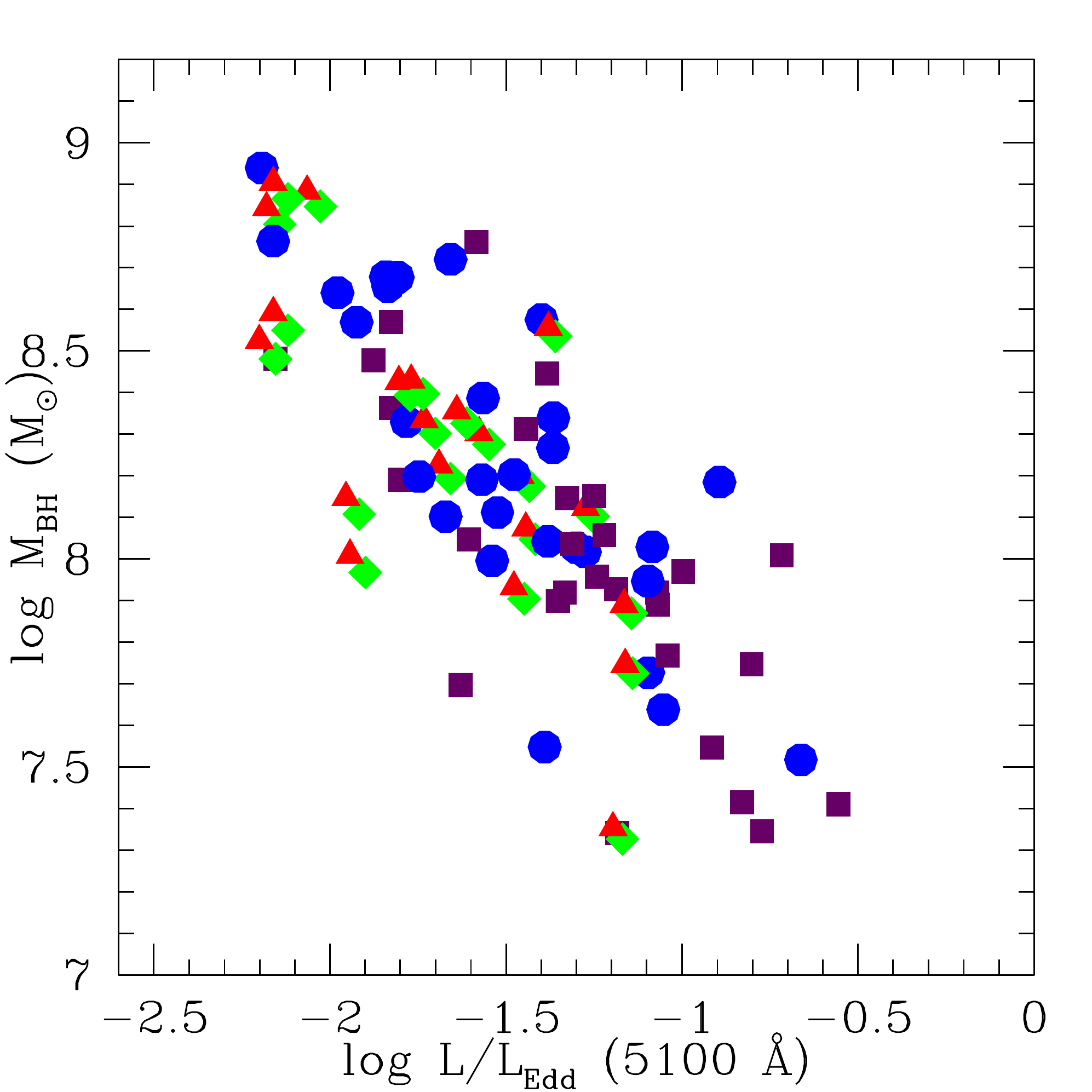}
\caption{The relation between $M_{BH}$ and $L/L_{Edd}$(5100 \AA) based on measurements from Mg II (purple squares), H$\beta$ (blue circles), H$\alpha$ (green diamonds) and H$\beta$ predicted from H$\alpha$ (red triangles). The inverse correlation is likely the result of two effects: cosmic downsizing and our luminosity cut. \\ (A color version of this figure is available in the online journal.) \label{fig:MbhvLfrac} }
\end{figure}

   Figure~\ref{fig:MbhvLfrac} shows our strongest correlation among AGN properties. The $M_{BH}$ increases as $L/L_{Edd}$ decreases, which probably arises as a result of two effects. The first effect is the dearth in massive BHs with high accretion rates at this redshift, which is the result of cosmic downsizing \citep{heckman04}. The second cause is the lack of lower mass black holes with low accretion rates which fail to make the luminosity cut. The combination of these two effects, demographics plus selection effects, leads to a strong inverse correlation. However this does not rule out the possibility that there is an underlying physical correlation, but confirming this would require a more sophisticated sample selection. 
   

\subsubsection{Narrow-Line Ratios} 

Narrow emission lines can be powered by several sources of photoionizing radiation, including young O and B stars and the central AGN. Comparing the relative strengths between lines with high and low-ionization potentials tells us about the shape of the photoionizing continuum and thus diagnoses the source. AGN have harder continua (relatively more high-energy photons) while star-forming continua are softer (relatively more low-energy photons). 

\begin{figure}[tbhp]
\figurenum{4} 
\centering
\includegraphics[scale=0.6]{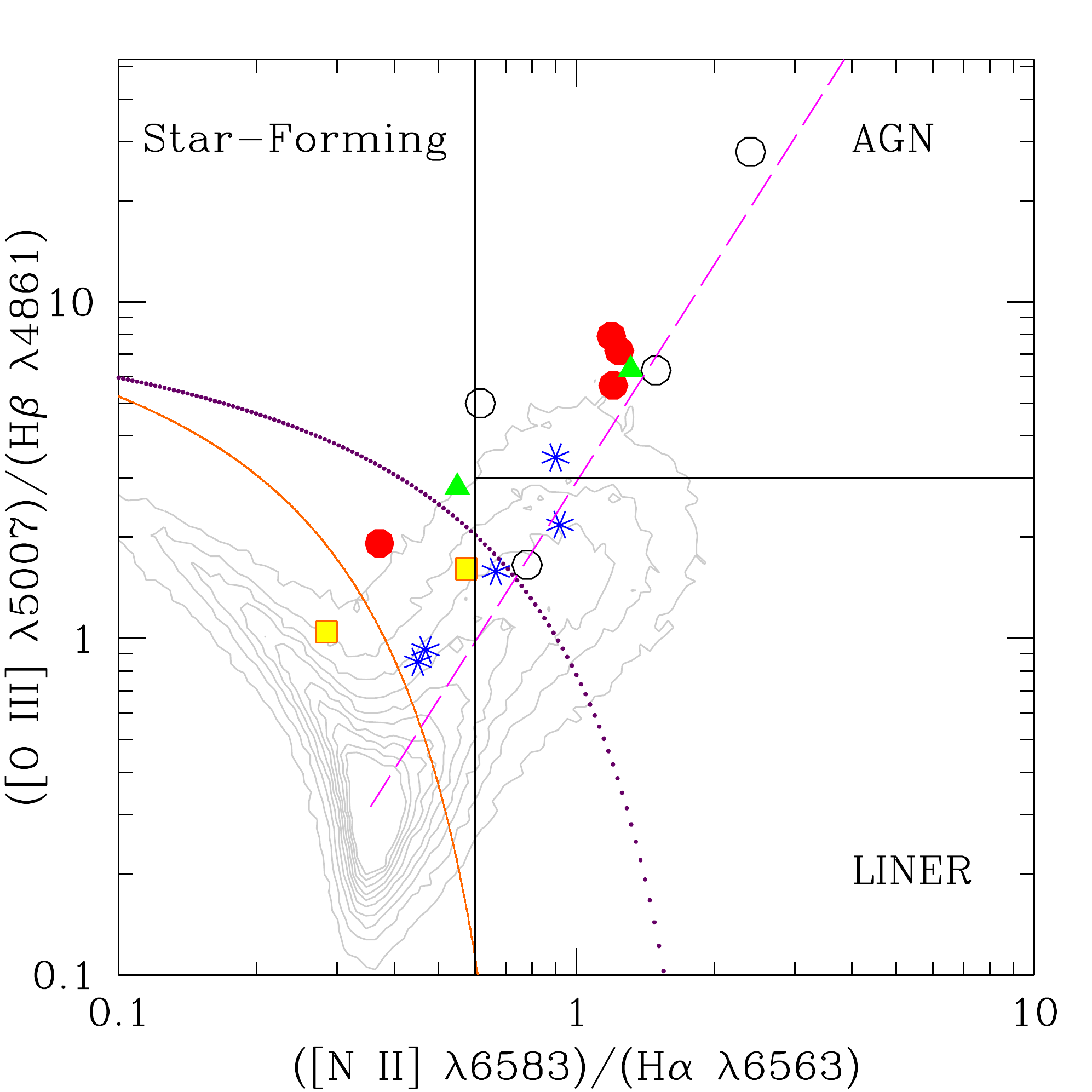}
\caption{Narrow line BPT diagnostic diagram. The point type indicates morphology: early-type (red circles), spiral (blue stars), ``probable'' spiral (green triangles), indeterminate morphology (yellow squares), and no morphological data (black open circles). The contours mark the distribution of SDSS DR7 galaxies \citep[after ][]{kauffmann03}. With black solid lines we make the traditional demarkations  of (1) star-formation if [\ion{N}{2}]/H$\alpha < 0.6$, (2) AGN for [\ion{N}{2}]/H$\alpha < 0.6$ and [\ion{O}{3}]/H$\beta > 3$, and, (3) LINER (low-ionization nuclear emission region) when [\ion{N}{2}]/H$\alpha < 0.6$ and [\ion{O}{3}]/H$\beta < 3$. The purple dotted curve gives the theoretical upper limit to the narrow-line ratios for star-forming regions given by \citet{kewley01}. The orange solid curve gives a more conservative empirical relation from \citet{kauffmann03} which marks the lower boundary between star-forming galaxies and star-forming plus AGN (composite-type) galaxies. Furthermore, the magenta dashed diagonal line beginning at the locus of galaxies ([\ion{N}{2}]/H$\alpha = -0.45$ and [\ion{O}{3}]/H$\beta = -0.5$) and extending towards the positive [\ion{O}{3}]/H$\beta$ axis and clockwise at an angle of $\phi = 25^{\circ}$ marks a mixing line. Starting from the locus and increasing in distance away along the mixing line, AGN become increasingly dominant. All PSQs show AGN activity in their spectra; however, PSQs fall along nearly the full extent of the mixing line indicating a  broad range in relative contributions from star-formation. \\ (A color version of this figure is available in the online journal.) \label{fig:BPT} }
\end{figure}

  We find a number of strong correlations among narrow-line ratios that can be interpreted in terms of the relative contributions of AGN and star-formation. In particular the highly ionized [\ion{O}{3}] $\lambda5007$ emission line is more prominent in AGN spectra but other lines may also be diagnostic. For example, Figure~\ref{fig:BPT} shows the traditional \citet*[][BPT]{baldwinphillipsterlevich81} diagram, [\ion{O}{3}]/H$\beta$ versus [\ion{N}{2}]/H$\alpha$ flux ratios. Diagrams such as this one help us understand the nature of relative contributions of ionizing radiation from different sources. PSQs fall along and above nearly the full extent of the mixing line, which indicates the presence of an AGN but a wide range in the relative contribution of star-formation.


\subsubsection{Starburst Properties} 

\begin{figure}[tbhp]
\figurenum{5} 
\centering
\includegraphics[scale=0.6]{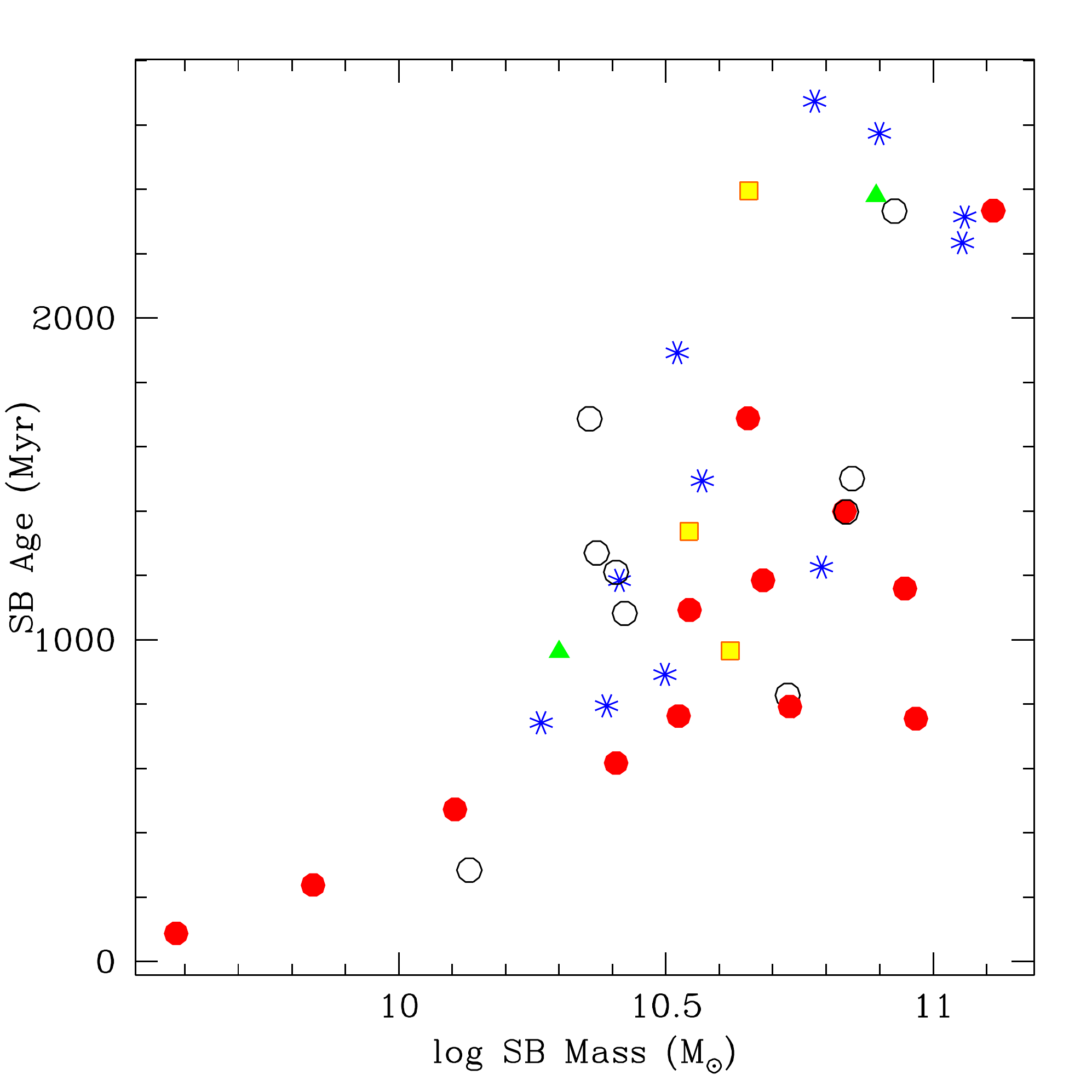}
\caption{Starburst mass versus age. The point type indicates morphology: early-type (red circles), spiral (blue stars), ``probable'' spiral (green triangles), indeterminate morphology (yellow squares), and no morphological data (black open circles). The dearth of points at low mass and old ages is due to the luminosity cut selection effect. However, the missing objects with young, massive starburst is unexpected. Thus, another class of objects such ULIRGs or post-starburst galaxies may be the parent population. \\ (A color version of this figure is available in the online journal.) \label{fig:SBplot} }
\end{figure}

   The strongest correlation among starburst properties is between the age and mass of the post-starburst population (Figure~\ref{fig:SBplot}). Just as in the case of AGN properties \S~\ref{sec:Anal.CorrAGN}, this correlation likely arises from a combination of demographics and selection effects. There is a lack of objects at large mass and young age. Such objects must exist and are likely dust enshrouded, seen in the infrared as luminous infrared galaxies \citep[LIRGs,][]{sandersmirabel96} or may be found as post-starburst galaxies if there is a significant time delay before the onset of AGN activity (see e.g., \citealt*{wild10} and \citealt{schawinski09}). The envelope in the upper left is the result of a selection effect: our luminosity limit.

\subsubsection{AGN versus Starburst Properties}

   We find that higher spectral indices (bluer AGN) correlate with larger starburst masses. This could be a result of a degeneracy between the spectral index and scaling of the starburst. As the spectral index increases it has the effect of taking away light, in which case the scaling of the starburst must make up the difference. Furthermore, if we assume a spectral index more typical of AGN we might expect a stronger correlation \citep[e.g.,][find find $\alpha\ \sim1.5$ with 0.5 scatter versus our 1.0]{francis93}. However, while the starburst template continuum guides our fitting, the Balmer region (i.e., Balmer break and absorption lines) of the spectra governs the quality of the starburst fit (see Figure~\ref{fig:zoom}). 

   Perhaps the most interesting result from this analysis is the lack of significant correlations between parameters which might have been suggested by merger-induced evolutionary scenarios \citep[e.g.,][]{dimatteo05,springel05,hopkins06}. Specifically, AGN luminosity (and/or Eddington ratio) should decline following a merger-triggered fueling event as the age of the starburst population increases. If there is a single mechanism driving both AGN and starburst activity, naturally leading to correlations between their properties, then our results present a problem for such models. Our selection effects that limit parameter space could limit our ability to see correlations, or there could be a delay in AGN triggering or  multiple types of triggering events. For the latter reason, we also perform analysis by morphological classification.


\begin{deluxetable}{lrrccrrccrrcrrc} 
\tabletypesize{\tiny}
\tablecolumns{15}
\tablewidth{0pc}
\tablecaption{Population Tests \label{tab:pop}}
\tablehead{
\colhead{ } & \multicolumn{3}{c}{Total} & \colhead {} & \multicolumn{3}{c}{Early-type} & \colhead {} &\multicolumn{3}{c}{Spirals} & \colhead{ } & \colhead{ } \\
\cline{2-4} \cline{6-8} \cline{10-12}
\colhead{Property} &        \colhead{ $\mu$\tablenotemark{a}} &      \colhead{$\sigma$\tablenotemark{b}} &        \colhead{Number\tablenotemark{c}} & \colhead{ } &         \colhead{$\mu$} &      \colhead{$\sigma$} &        \colhead{Number} & \colhead{ } &         \colhead{$\mu$} &      \colhead{$\sigma$} &       \colhead{Number} &           \colhead{$W$\tablenotemark{d}} &      \colhead{P (\%)} }
\startdata
$z$ &          0.32 &          0.05 &            38 & &          0.33 &          0.05 &            13 & &          0.31 &          0.05 &            13 &      \ldots\ldots &      \ldots\ldots \\
log $\lambda L_{\lambda}$(5100 \AA) &         43.94 &          0.26 &            38 & &         44.11 &          0.23 &            13 & &         43.77 &          0.24 &            13 &        -3.051 &         0.114 \\
$\alpha$ &          1.00 &          0.66 &            38 & &          1.11 &          0.66 &            13 & &          1.06 &          0.70 &            13 &      \ldots\ldots &      \ldots\ldots \\
$f_{nuc}$ &          0.55 &          0.13 &            38 & &          0.57 &          0.15 &            13 & &          0.51 &          0.12 &            13 &      \ldots\ldots &      \ldots\ldots \\
log $L_{Tot}$ &         43.93 &          0.20 &            38 & &         44.08 &          0.18 &            13 & &         43.81 &          0.16 &            13 &        -3.359 &         0.039 \\
log $L_{AGN}$ &         43.65 &          0.24 &            38 & &         43.82 &          0.23 &            13 & &         43.51 &          0.22 &            13 &        -3.205 &         0.068 \\
log $L_{SB}$ &         43.57 &          0.20 &            38 & &         43.69 &          0.20 &            13 & &         43.50 &          0.14 &            13 &        -2.590 &         0.480 \\
log $M_{BH}$ Mg II &          7.99 &          0.37 &            29 & &          8.04 &          0.40 &            12 & &          8.08 &          0.39 &             9 &      \ldots\ldots &      \ldots\ldots \\
$L/L_{Edd}$ Mg II &          0.07 &          0.06 &            29 & &          0.08 &          0.06 &            12 & &          0.05 &          0.05 &             9 &      \ldots\ldots &      \ldots\ldots \\
log $M_{BH}$ H$\beta$ &          8.27 &          0.41 &            30 & &          8.28 &          0.39 &            13 & &          8.49 &          0.45 &             7 &      \ldots\ldots &      \ldots\ldots \\
$L/L_{Edd}$ H$\beta$ &          0.04 &          0.04 &            30 & &          0.05 &          0.04 &            13 & &          0.02 &          0.02 &             7 &      \ldots\ldots &      \ldots\ldots \\
log $M_{BH}$ H$\alpha$ &          8.25 &          0.38 &            21 & &          8.18 &          0.29 &             6 & &          8.30 &          0.38 &             9 &      \ldots\ldots &      \ldots\ldots \\
$L/L_{Edd}$ H$\alpha$ &          0.03 &          0.02 &            21 & &          0.04 &          0.02 &             6 & &          0.02 &          0.02 &             9 &      \ldots\ldots &      \ldots\ldots \\
log $M_{BH}$ H$\alpha$-H$\beta$ &          8.28 &          0.38 &            21 & &          8.21 &          0.29 &             6 & &          8.34 &          0.39 &             9 &      \ldots\ldots &      \ldots\ldots \\
$L/L_{Edd}$ H$\alpha$-H$\beta$ &          0.03 &          0.02 &            21 & &          0.04 &          0.02 &             6 & &          0.02 &          0.02 &             9 &      \ldots\ldots &      \ldots\ldots \\
log $M_{BH}$ Adopted &          8.17 &          0.36 &            37 & &          8.19 &          0.33 &            13 & &          8.20 &          0.38 &            13 &      \ldots\ldots &      \ldots\ldots \\
$L/L_{Edd}$ Adopted &          0.05 &          0.05 &            37 & &          0.06 &          0.04 &            13 & &          0.03 &          0.04 &            13 &      \ldots\ldots &      \ldots\ldots \\
log SB Mass &         10.58 &          0.33 &            38 & &         10.53 &          0.45 &            13 & &         10.65 &          0.28 &            13 &      \ldots\ldots &      \ldots\ldots \\
SB Age &       1321.75 &        689.74 &            38 & &        967.60 &        609.97 &            13 & &       1643.17 &        724.48 &            13 &         2.333 &         0.982 \\
$$[O III]/H$\beta$ &          4.83 &          5.29 &            32 & &          6.56 &          4.05 &            10 & &          2.54 &          1.59 &            11 &        -3.053 &         0.113 \\
$$[N II]/H$\alpha$ &          1.06 &          0.92 &            21 & &          1.57 &          1.47 &             6 & &          0.73 &          0.27 &             9 &      \ldots\ldots &      \ldots\ldots \\
$$[O III]/[O II] &          2.31 &          1.60 &            33 & &          2.79 &          1.60 &            11 & &          1.59 &          0.77 &            11 &      \ldots\ldots &      \ldots\ldots \\
$$[Ne V]/[Ne III] &          1.35 &          2.04 &            33 & &          2.07 &          2.13 &            11 & &          1.61 &          0.96 &            11 &      \ldots\ldots &      \ldots\ldots \\
$$Fe II/[O III] &          0.01 &          0.02 &            12 & &          0.01 &          0.02 &             6 & &          0.03 &          0.03 &             2 &      \ldots\ldots &      \ldots\ldots \\
\enddata
\tablenotetext{a}{The mean of the population.}
\tablenotetext{b}{The standard deviation of the population.}
\tablenotetext{c}{The number of objects that were used to calculate mean, standard deviation and/or population statistics.}
\tablenotetext{d}{When statistical differences in sample means exist at below the 1\% level we include the nonparametric Mann-Whitney statistic and its associated P-Value.}
\end{deluxetable}
\clearpage

\subsection{Dependence on Morphological Class} 
\label{sec:Anal.Pop}

\begin{figure}[!h]
\figurenum{6} 
  \centering
   \subfloat[][(a) \label{fig:hist_lgtotl}]{\includegraphics[scale=0.3]{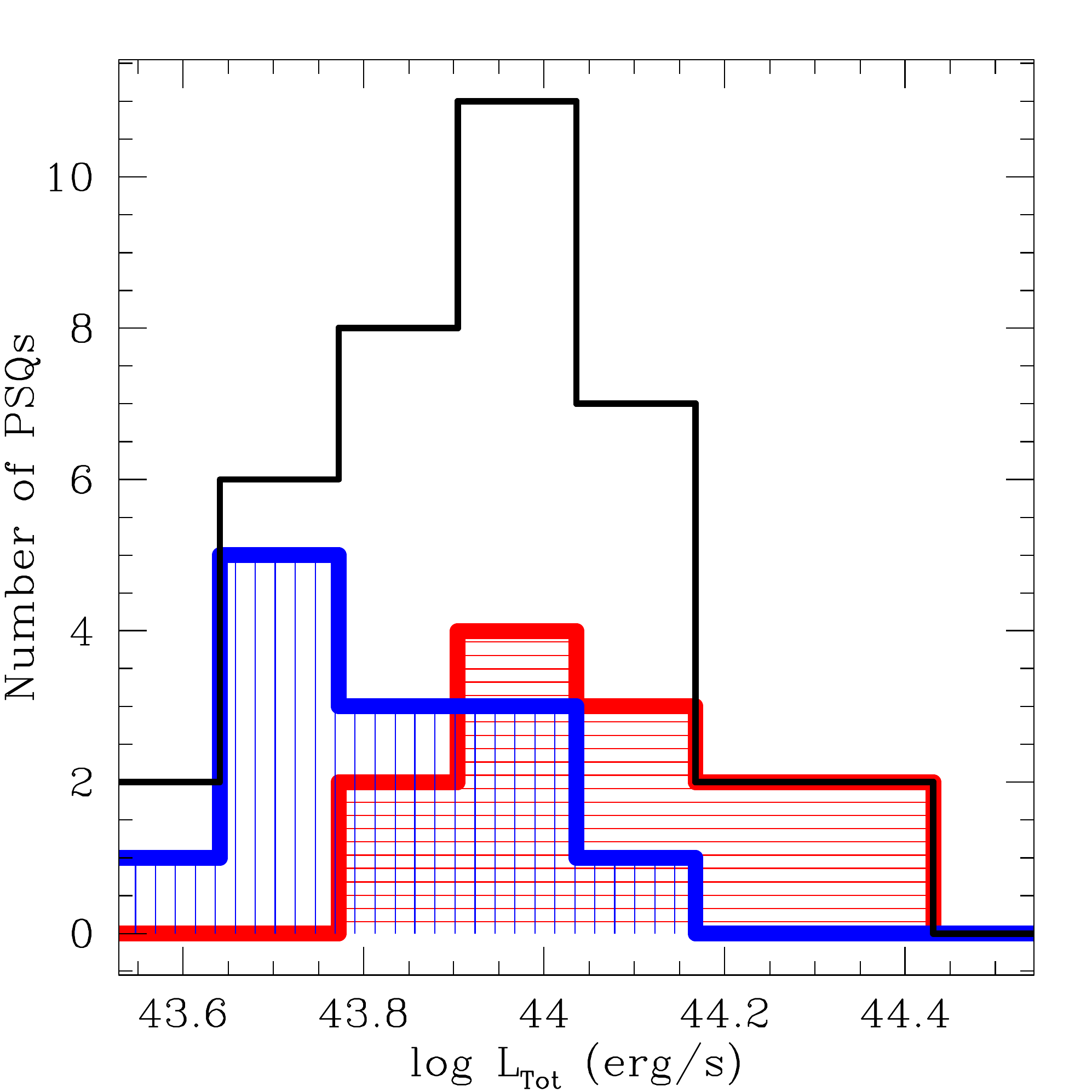}}
   \subfloat[][(b) \label{fig:hist_lglagn}]{\includegraphics[scale=0.3]{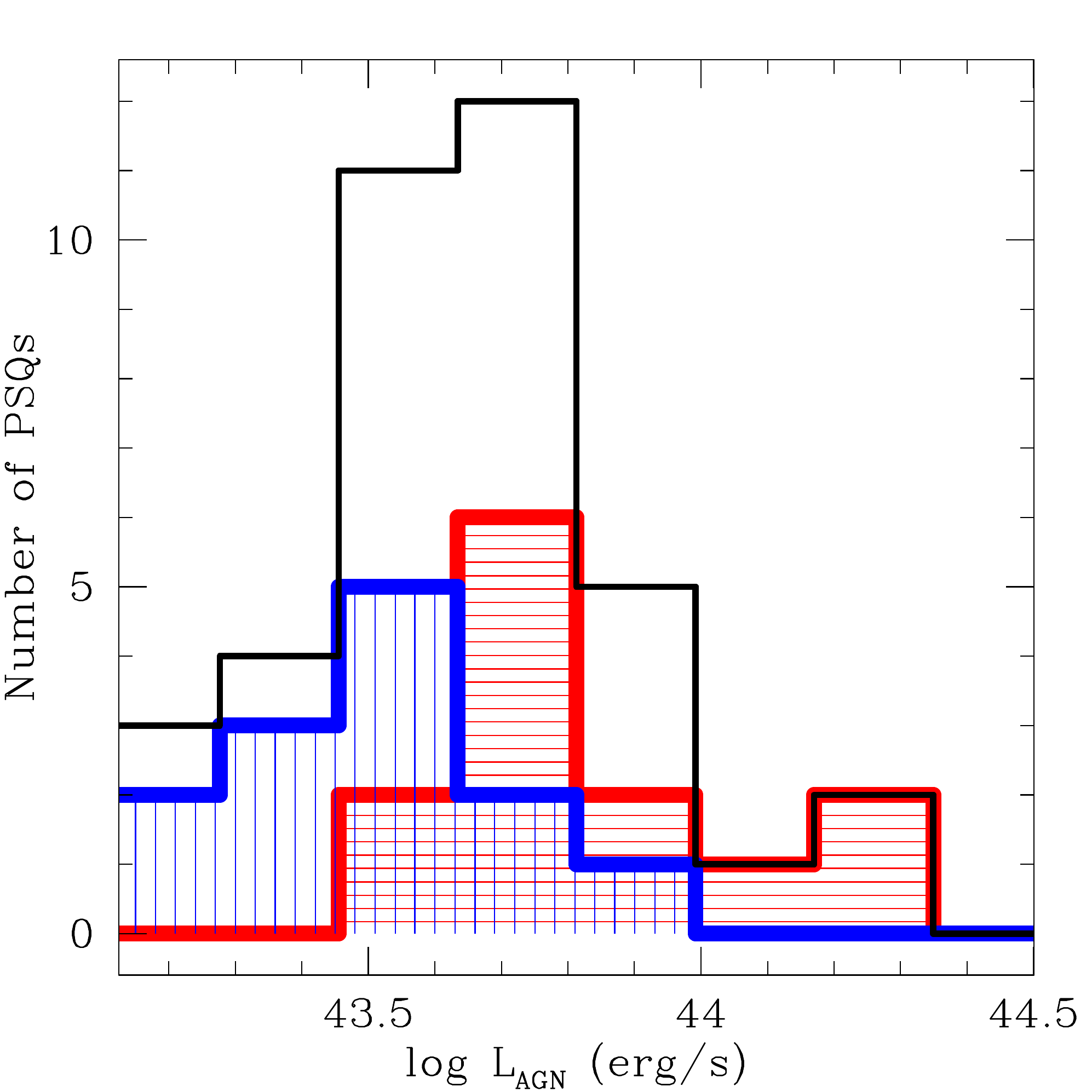}}
   \subfloat[][(c) \label{fig:hist_lglsb}]{\includegraphics[scale=0.3]{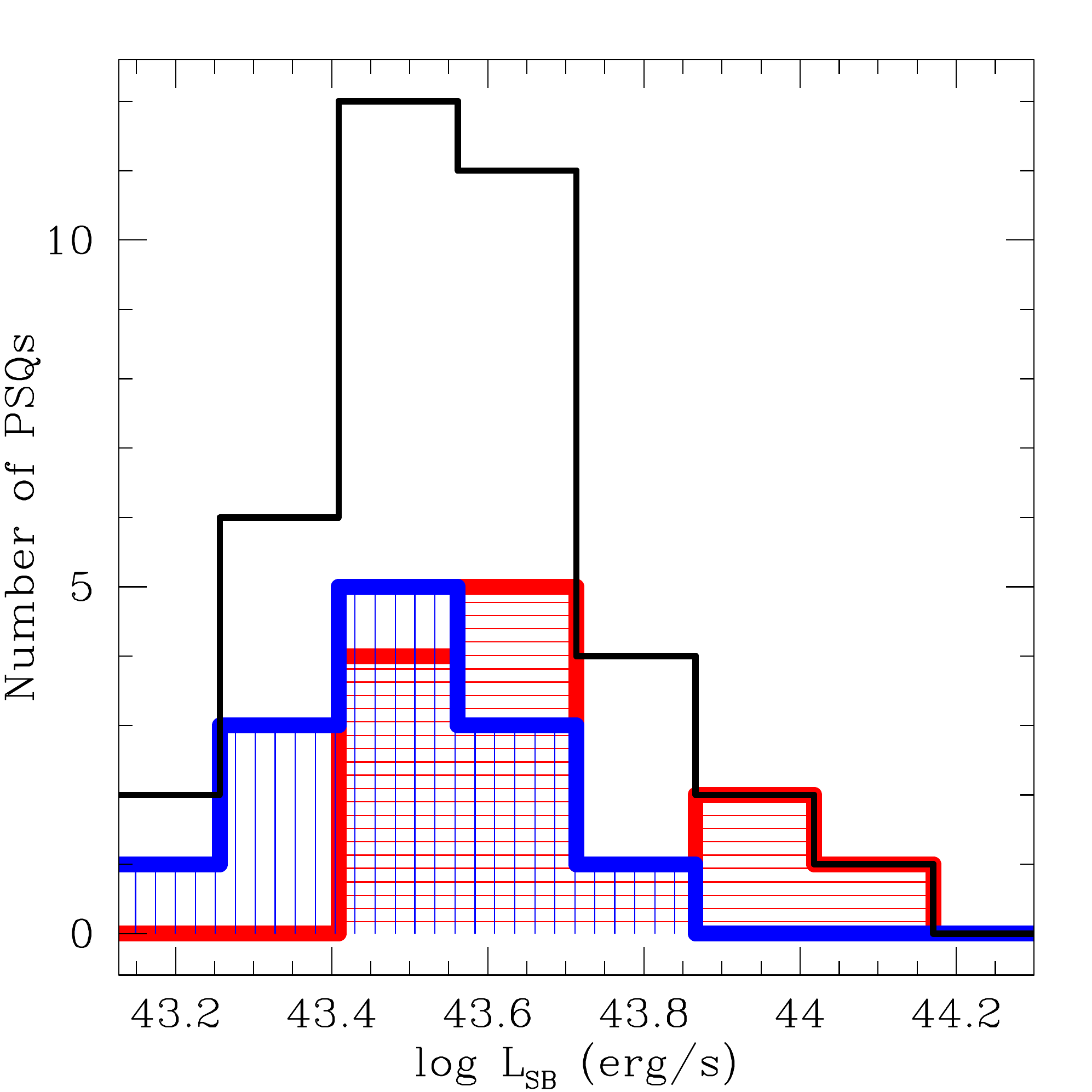}}\\
   \subfloat[][(d) \label{fig:hist_adopt_lfrac}]{\includegraphics[scale=0.3]{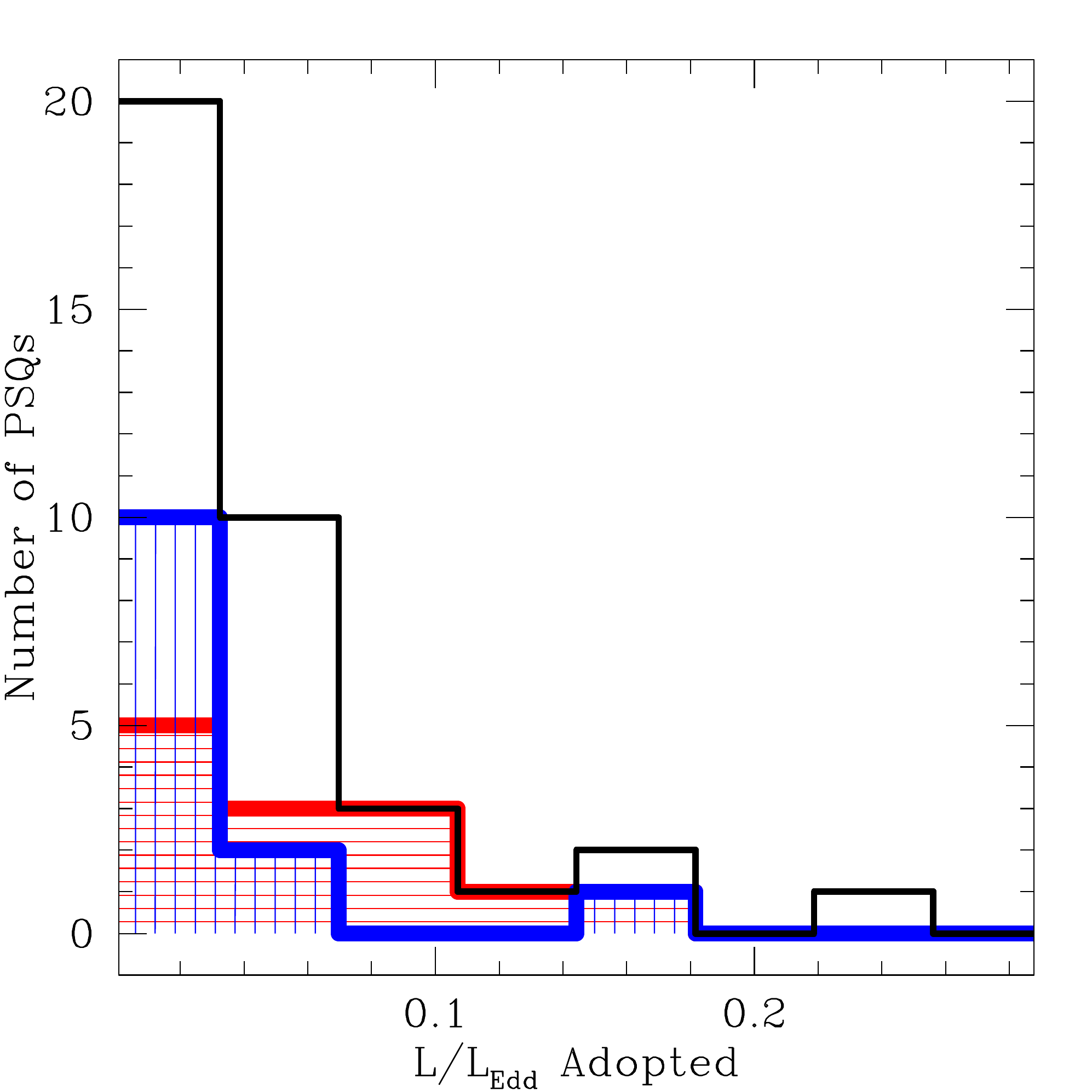}}
   \subfloat[][(e) \label{fig:hist_adopt_lgmbh}]{\includegraphics[scale=0.3]{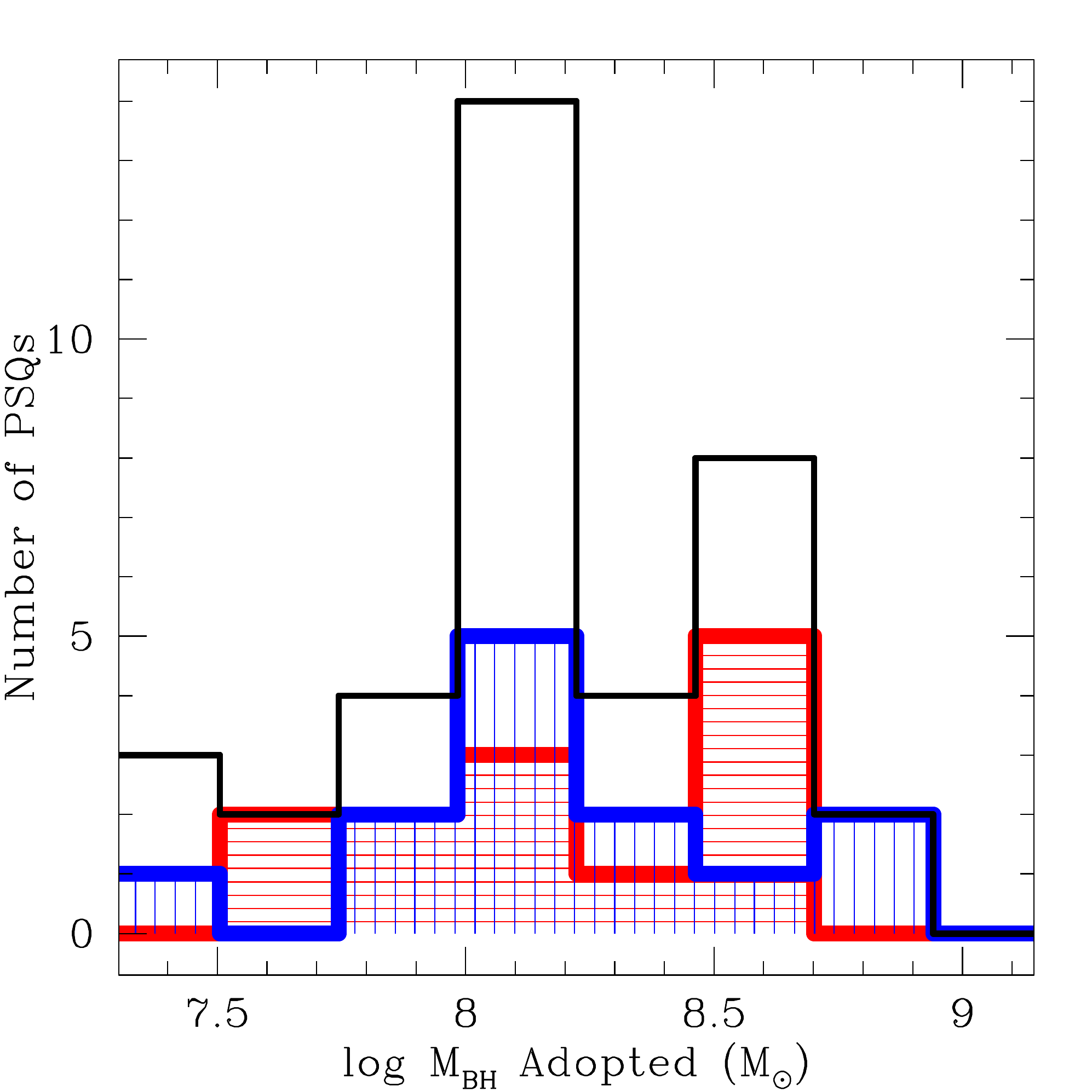}}
   \subfloat[][(f) \label{fig:hist_lgsbmass}]{\includegraphics[scale=0.3]{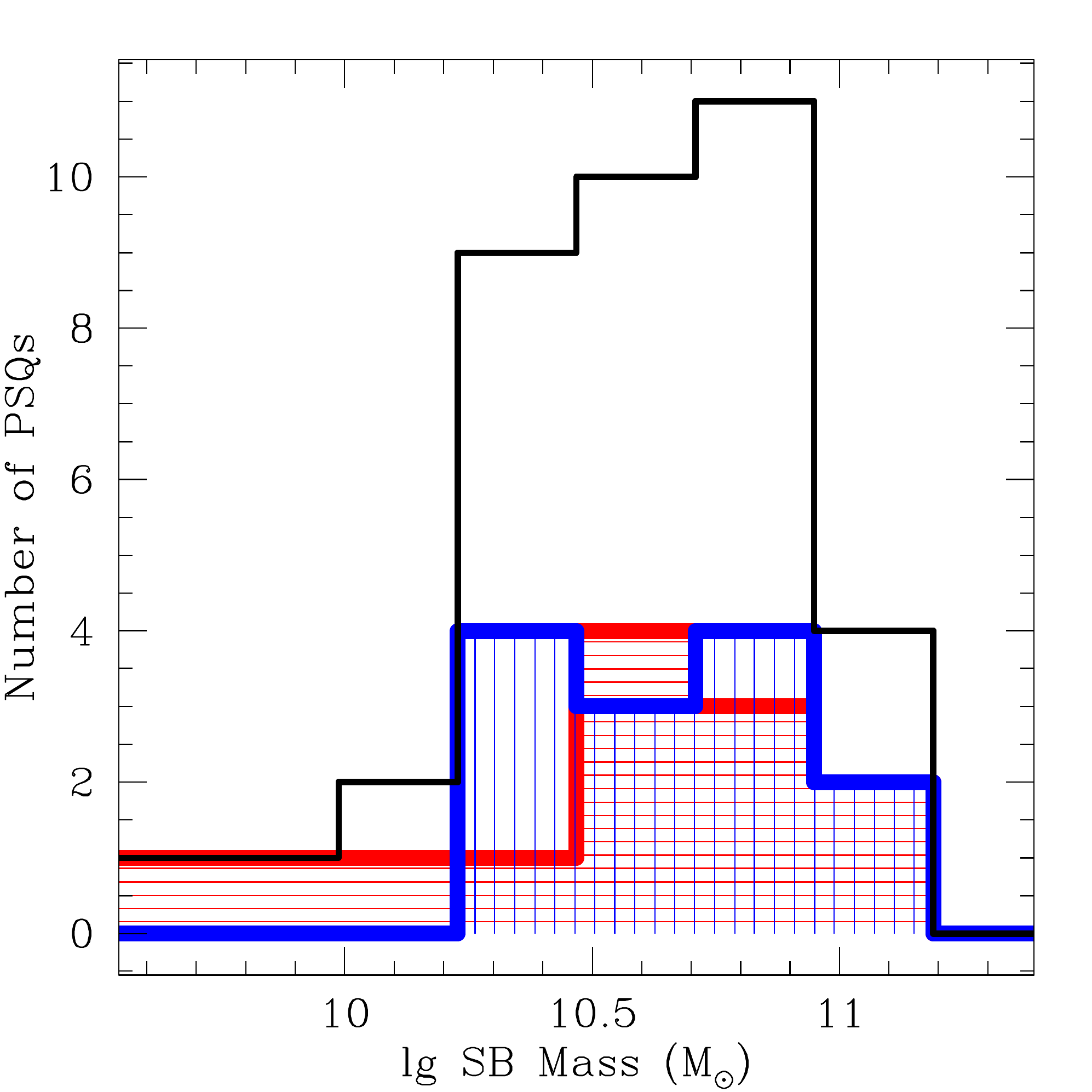}}\\
   \subfloat[][(g) \label{fig:hist_oiii2hb}]{\includegraphics[scale=0.3]{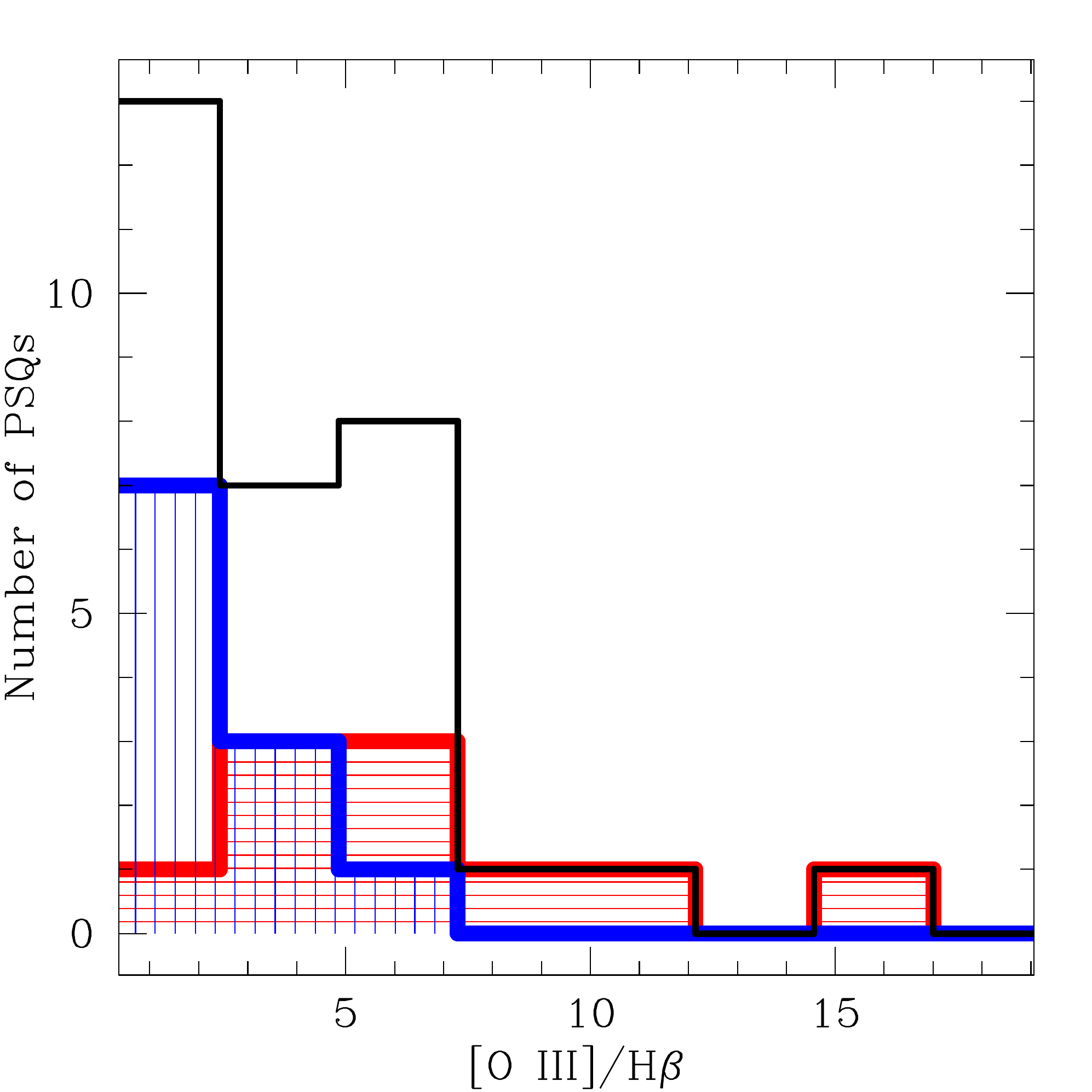}}    
   \subfloat[][(h) \label{fig:hist_sbage}]{\includegraphics[scale=0.3]{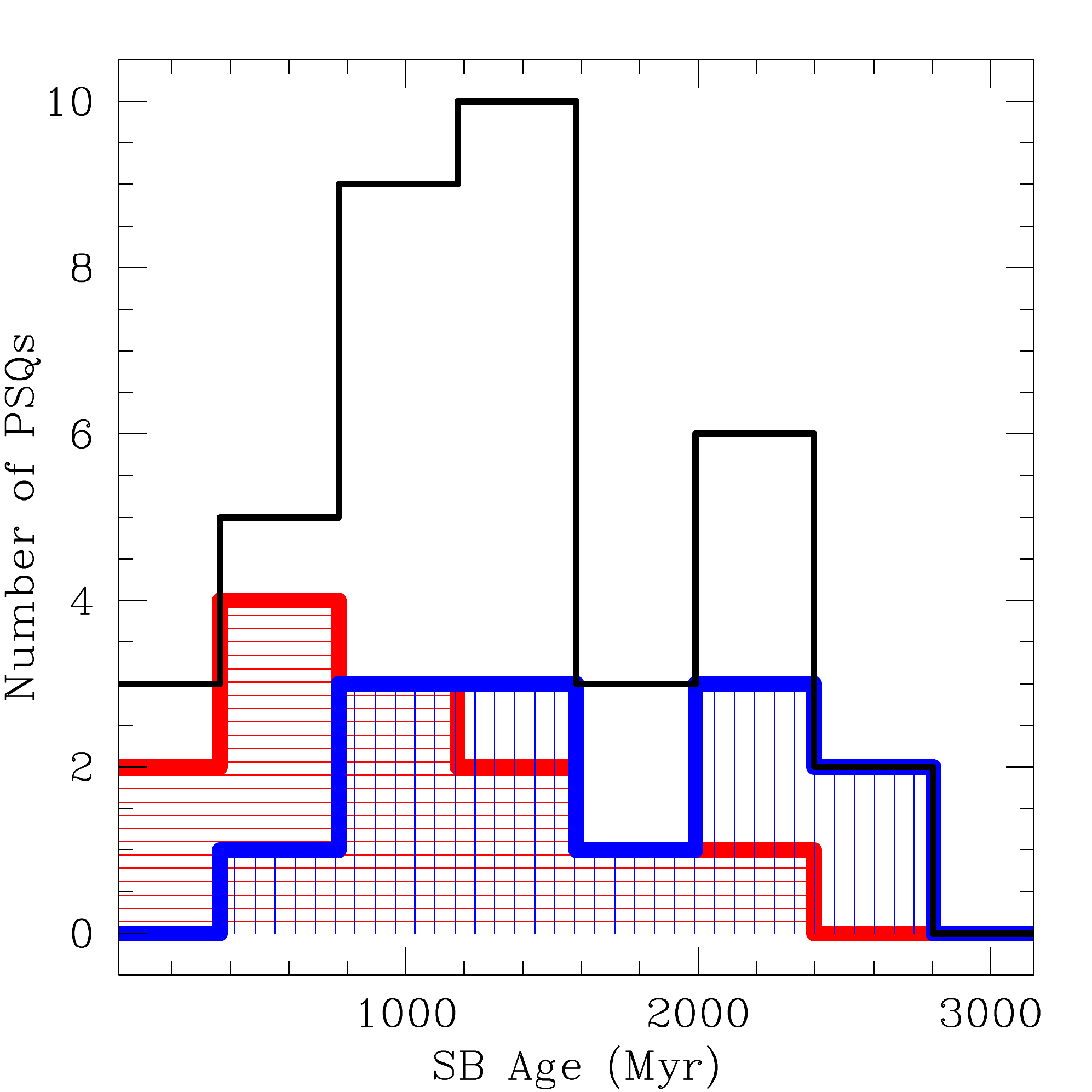}}\\
   \caption{\small{(a-h) The solid histogram represents the total distribution, while the red horizontal and blue vertical hatched histograms indicate the early-type and spiral plus ``probable'' spiral morphology distributions, respectively. Early-type PSQs have more luminous AGN and younger starburst populations, while PSQs with spiral hosts are consistent with ongoing star formation. \\ (A color version of this figure is available in the online journal.) \label{fig:morphhist} }}
  \centering
\end{figure}

  We investigate the relationships among morphology and fundamental (fitted and derived) properties of the AGN, narrow-line emission and starburst. Visual classifications were performed by three of the authors and the results are presented in \citet{cales11}. Generally, arms and bars distinguish spiral galaxies while smooth somewhat featureless (i.e., lacking arms/bars) light distributions identify early-type galaxies. We note that the early-type hosts are allowed to show tidal features although sometimes disturbances make classification difficult resulting in an indeterminate classification. In Table~\ref{tab:pop} we list the mean, standard deviation, and number of objects for our basic properties for the total sample, early-type and spiral (including ``probable'' spirals) subpopulations. When reliable statistical differences in sample means exist at P-Values below the 1\% level (better than $2.58\sigma$) we include the nonparametric Mann-Whitney statistic and its associated P-Value and give their distributions in Figure~\ref{fig:morphhist}. For reference, the mean, standard deviation and number of objects for the total sample are also given.
  
  Due to the small number of objects in the subpopulations and the lack of significant differences below the 1\% level between the disturbed and undisturbed populations we do not present population statistics for classifications of this type.

The significant differences for fundamental parameters within the morphology subclasses are:
\begin{enumerate}
\item \textbf{AGN:} The PSQs hosted in early-type galaxies have statistically significant higher total luminosities than spiral hosted PSQs, and this appears to be somewhat more driven by differences in AGN luminosity although starburst luminosities also differ. This is supported by various methods of measuring the PSQ luminosity and its components (i.e., $L_{Tot}$, $L_{AGN}$ and $L_{SB}$, see Figures~\ref{fig:morphhist}a-c). We note that AGN luminosity increases with black hole mass and/or accretion rate. While there are no statistically significant differences between the subpopulations for Eddington fraction, black hole mass or starburst mass, it is possible that the AGN with early-type hosts tend to have higher luminosities because they have higher Eddington ratios (Figures~\ref{fig:morphhist}d-f).

\item \textbf{NLR:} Figure~\ref{fig:morphhist}e shows that the PSQs hosted in early-type galaxies have larger [\ion{O}{3}]/H$\beta$ than spiral hosted PSQs, and we interpret this as differences in the hardness of the ionizing continuum reflecting the relative contributions of the AGN and star-formation. The spiral PSQs have more on-going star-formation.

\item \textbf{Starburst:} The PSQs hosted in early-type galaxies tend to have younger starburst ages from our fitting results than the PSQs hosted by spirals (Figure~\ref{fig:morphhist}f). This may be interpreted in different ways. Some possible explanations of this are: (1) the starburst luminosities are smaller in the spiral hosted PSQs and our single age fits may be skewed by the presence of an older stellar population, (2) chance snapshot of unassociated starburst and  AGN activity in spiral hosts with large starburst ages, or (3) longer timescales between triggering of the AGN and starburst in spiral PSQs. There may be no clear interpretation of this effect without additional data.


\end {enumerate}
   


\clearpage




\subsection{Spectral Diagnostics for Mode of Growth} 
\label{sec:Anal.Mode}

   PSQs appear to display both merger-driven and secular forms of galaxy evolution. They are a heterogeneous population hosted by both early-type and spiral galaxies, some of which appear to be results of major mergers while others exist in isolated systems. C11 postulates that the early-type PSQs may be the low-z analogs of luminous merger-induced $z\sim2$ quasars. Furthermore, at least some of the spiral PSQs are likely undergoing secular evolution. This conclusion is consistent with current theoretical frameworks which argue for two fundamentally different fueling mechanisms  (i.e., merger induced versus secular) responsible for mutual SMBH-bulge growth at the characteristic dividing line of quasar-Seyfert luminosities \citep{hopkinshernquist09}.
      
\begin{figure}[!h]
\figurenum{7} 
  \centering
    \subfloat[][(a) \label{fig:BPT2a}]{\includegraphics[scale=0.4]{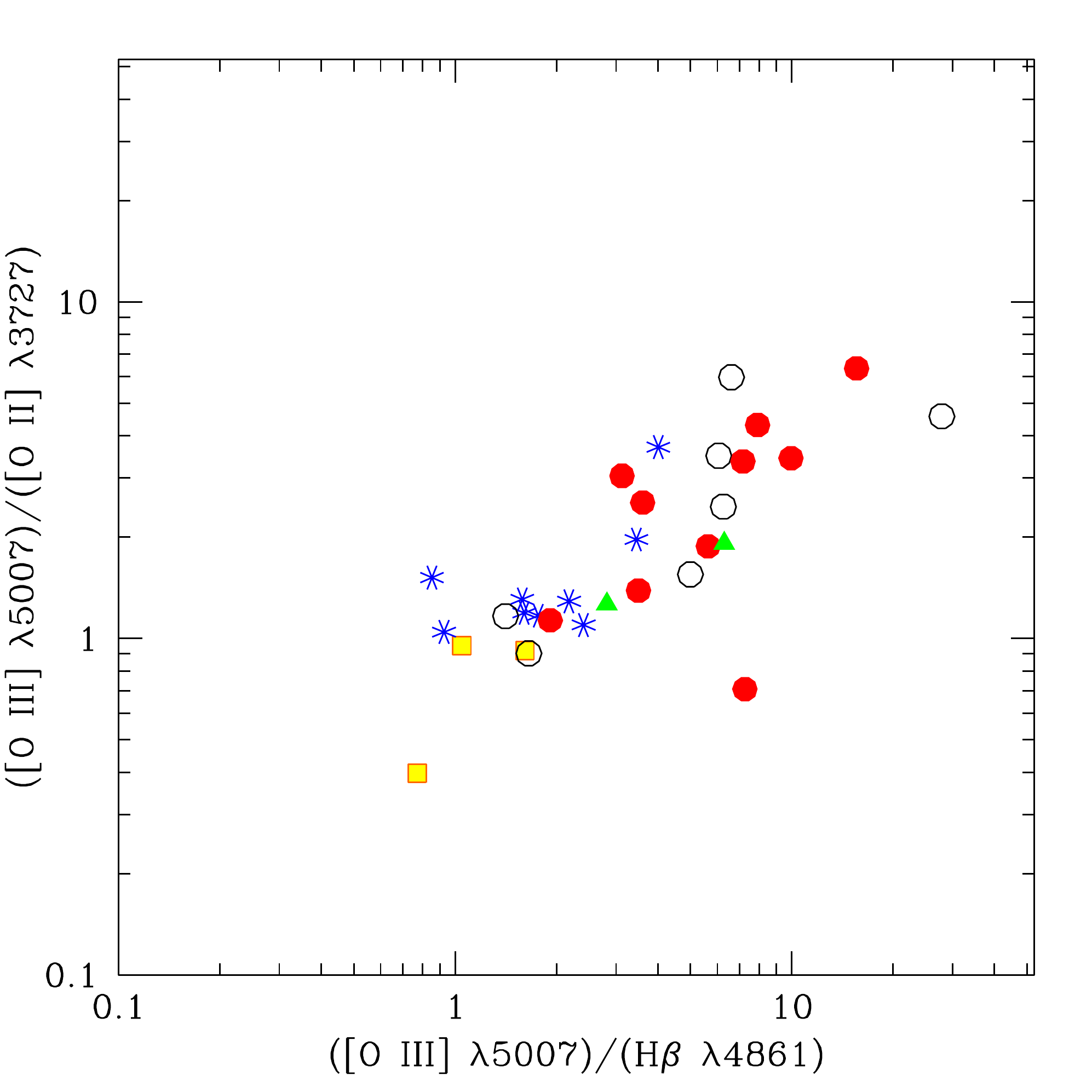}}
    \subfloat[][(b) \label{fig:BPT2b}]{\includegraphics[scale=0.4]{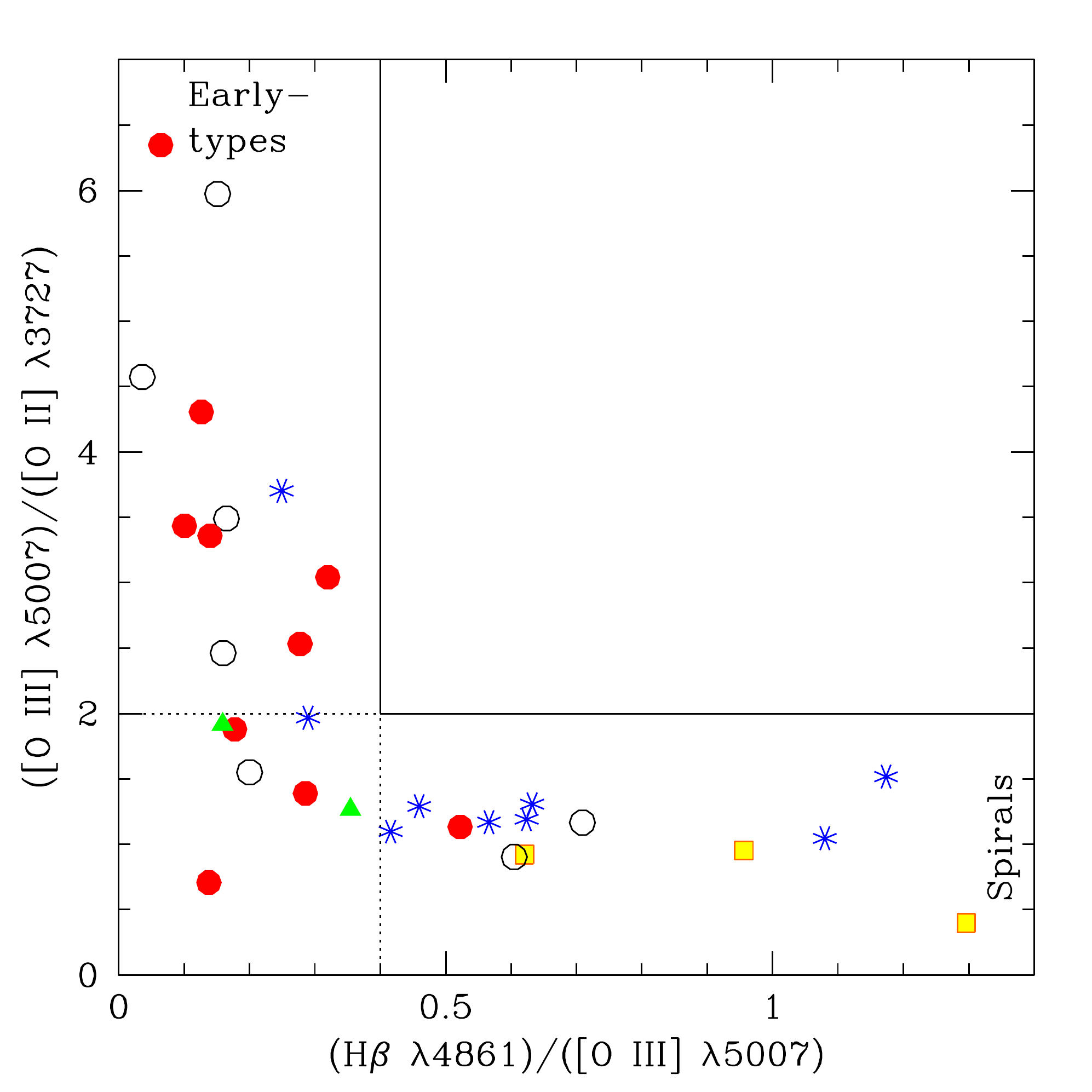}} \\
     \subfloat[][(c) \label{fig:BPT2c}]{\includegraphics[scale=0.4]{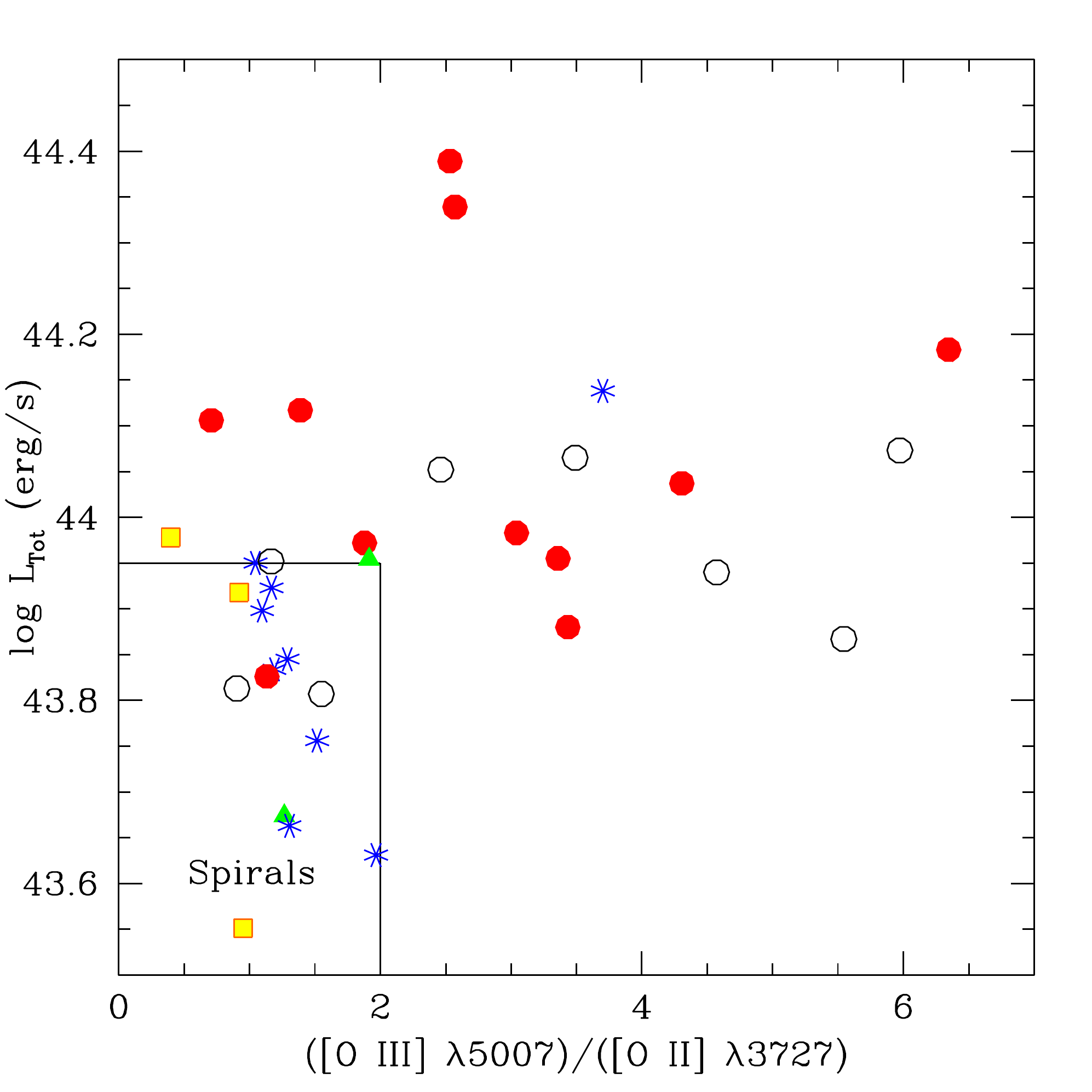}}
    \subfloat[][(d) \label{fig:BPT2d}]{\includegraphics[scale=0.4]{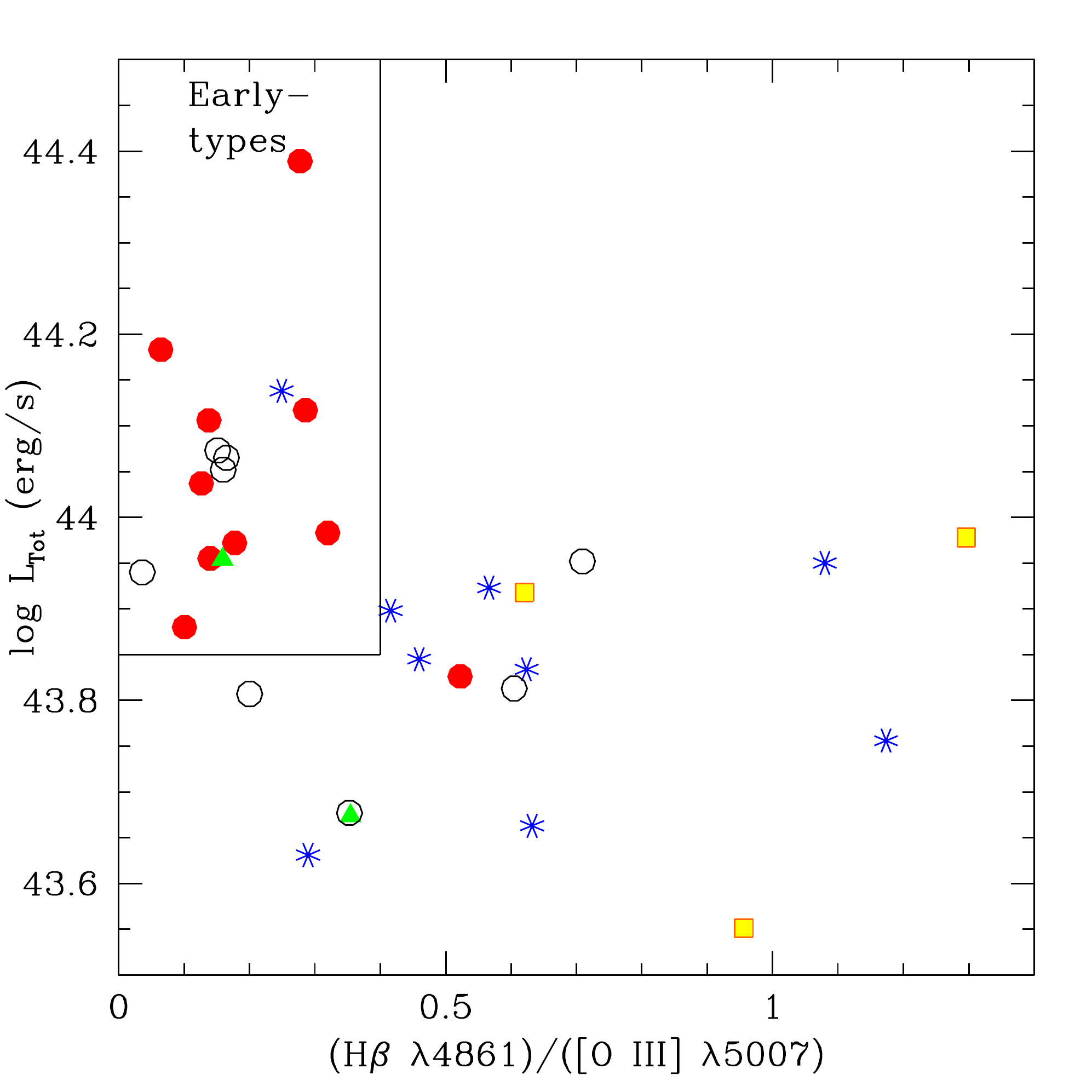}} \\
\caption{\small{Diagnostic diagrams featuring [\ion{O}{3}]/[\ion{O}{2}], [\ion{O}{3}]/H$\beta$, and $L_{Tot}$. The point type indicates morphology: early-type (red circles), spiral (blue stars), ``probable'' spiral (green triangles), indeterminate morphology (yellow squares), and no morphological data (black open circles). (a) The log-log [\ion{O}{3}]/[\ion{O}{2}] versus [\ion{O}{3}]/H$\beta$ plot. (b) Linear [\ion{O}{3}]/[\ion{O}{2}] versus H$\beta$/[\ion{O}{3}]. (c) $L_{Tot}$ versus [\ion{O}{3}]/[\ion{O}{2}]. (d) $L_{Tot}$ versus H$\beta$/[\ion{O}{3}]. PSQs with early-type and spiral hosts lie in very different regions of the three-dimensional space given by $L_{Tot}$, [\ion{O}{3}]/[\ion{O}{2}] and H$\beta$/[\ion{O}{3}]. We note where early-type and spiral PSQs lie according to our selection criteria.  \\ (A color version of this figure is available in the online journal.) \label{fig:BPT2} }}
  \centering
\end{figure}

We explored issues of the relative contributions of AGN and star-formation to powering narrow-emission lines previously in the context of the BPT diagram. We have also found differences between a number of properties and morphological subtype. We can combine this information to create observational selection criteria to select PSQs as a function of galaxy host type without requiring high spatial resolution imaging. 
With the exception of a rare outlier we can use a combination of H$\beta$/[\ion{O}{3}], [\ion{O}{3}]/[\ion{O}{2}], and $L_{Tot}$ to distinguish host galaxy type. Figure~\ref{fig:BPT2} shows several plots involving these quantities and where the objects of different classification fall. Using these we establish the following parameters to select PSQs by morphological subtype. 

  
\begin{itemize}
\item \textbf{Early-type:} L$_{Tot}$ $>$ 10$^{43.85}$ ergs s$^{-1}$ $+$ H$\beta$/[\ion{O}{3}] $<$ 0.4
\item \textbf{Spiral:} L$_{Tot}$ $<$ 10$^{43.95}$ ergs s$^{-1}$ $+$ [\ion{O}{3}]/[\ion{O}{2}] $<$ 2.0
\end{itemize}


   While it has been suggested that major-mergers are not important drivers of AGN activity in the local downsized universe \citep{cisternas11}, such objects do exist primarily among quasars and may be valuable to study as analogs of luminous high redshift quasars \citep{canalizo07,bennert08,bennert10}. Studies of local AGN ($z\sim0.05$) have also found two modes of growth distinguished by morphology. \citet{schawinski10a} find that the least massive SMBH population of early-type hosts are growing while the most massive SMBH population of late-type hosts are growing. Furthermore, \citet{schawinski10b} trace a correlation between merger fraction of early-type hosts and an evolutionary sequence involving starbursts and AGN, such that, even in the local universe, early-type galaxies are still merger-induced. This has important ramifications for future studies involving galaxy evolution and should guide the path of such studies.

\clearpage

\section{Summary}
\label{sec:Summ}

We performed AGN-host spectral decomposition of 38 PSQs, 29 of which have morphological classifications from C11. We characterized the host starburst masses and ages, and the AGN black hole masses and Eddington fractions, with the aim of improving our understanding of mutual evolution of luminous AGN and their hosts.
\begin{enumerate}
\item \textbf{Black Hole Properties:} The PSQs have $M_{BH} \sim10^{7.5-8.5}$ M$_{\odot}$ which are accreting at $\sim$1-10\% of Eddington luminosity.
\item \textbf{Starburst Properties:} The PSQs have massive starbursts $\sim$10$^{10-11}$ M$_{\odot}$ that span a range of ages $\sim 200-2000$ Myrs.
\item We find no strong correlations linking $M_{BH}$ and starburst properties.
\item The PSQ sample lacks young, massive starbursts that are likely obscured LIRGs that we cannot select optically or are seen as post-starburst galaxies.
\item \textbf{Narrow-line emission:} When plotted on traditional BPT diagrams, PSQs fall along and slightly above the mixing line showing a wide range in the relative contributions of AGN and star-formation.
\item \textbf{Morphology:} Early-type PSQs have significantly stronger AGN luminosities, younger ISB ages, and narrow-line ratios indicative of harder photoionizing continua when compared to spiral PSQs.
\item \textbf{Morphological Selection of PSQs:} We determine that the selection criterion for (1) early-type PSQs of L$_{Tot}$ $>$ 10$^{43.85}$ ergs s$^{-1}$ and H$\beta$/[\ion{O}{3}] $<$ 0.4 and (2) spiral PSQs of L$_{Tot}$ $<$ 10$^{43.95}$ ergs s$^{-1}$ and [\ion{O}{3}]/[\ion{O}{2}] $<$ 2.0 is efficient for determining morphology of the PSQ sample.
\item We conclude that the PSQ sample displays two distinct mechanisms for joint AGN and starburst activity. The higher luminosity early-type PSQs appear to be the product of major-mergers, show little current star formation and may be identified as the low-z analogs of the luminous and likely merger-induced high redshift quasars. PSQs hosted in spirals likely represent a lower luminosity mode of activity, such as, `Seyfert mode' or secular activity, triggered by internal processes, e.g., bars, or external triggers, e.g., harassment.
\end{enumerate}

In order to further explore possible correlations between AGN and starburst properties we intend to study a larger sample of PSQs with a broader range in AGN and starburst strengths. For example, plotting PSQs on the $M_{BH}-\sigma_{*}$ relationship may be illuminating in better understanding their role in massive galaxy evolution \citep[][\textit{submitted}]{hiner12}. Another way to do this is to explore lower luminosity, lower redshift objects of the \citet{brotherton07} catalog in conjunction with morphologies from Galaxy Zoo and spectral fitting of only the highest quality SDSS spectra.

\acknowledgements

We acknowledge support from NASA through the LTSA grant NNG05GE84G. Z. Shang acknowledges support from the national Natural Science Foundation of China through grant 10633040 and support by Chinese 973 Program 2007CB815405. G. C. acknowledges support from the National Science Foundation, under grant number AST 0507450.  S. L. Cales was supported in part by NASA Headquarters under the NASA Earth and Space Science Fellowship Program (Grant NNX08AX07H), in part by National Science Foundation GK-12 Program (Project 0841298) and also in part by ALMA-CONICYT program 31110020.

\bibliographystyle{/Applications/TeX/BibLib/apj}
\bibliography{/Applications/TeX/BibLib/AllCat}
\clearpage

\section*{Appendix - Spectra of Fitting Results }
\label{sec:App}

Spectral decomposition of AGN and post-starburst stellar population. The red line is the data. The blue lines make up the components used in the fitting. For ease of interpretation we plot the emission lines and \ion{Fe}{2} templates above the power-law. The black line is the model fit to the data.

\begin{figure}[tbhp]
   \centering
   \figurenum{8} 
      \subfloat[][Fig. 8$a$]{\includegraphics[width=4.1in]{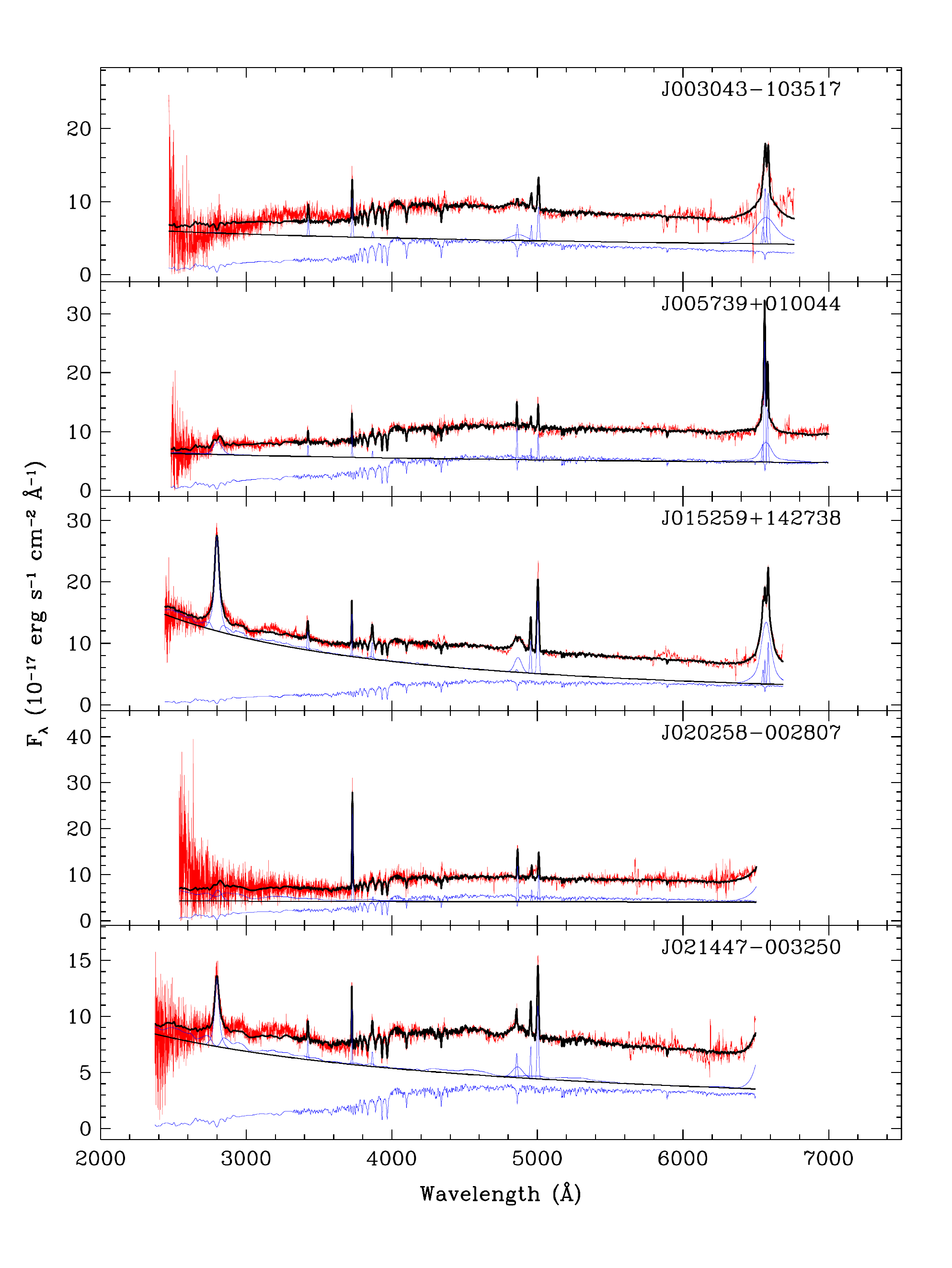}}
      \caption{A color version of this figure is available in the online journal. \label{fig:modspec} }
\end{figure}
\begin{figure}
   \ContinuedFloat
   \centering
      \subfloat[][Fig. 8$b$]{\includegraphics[width=6in]{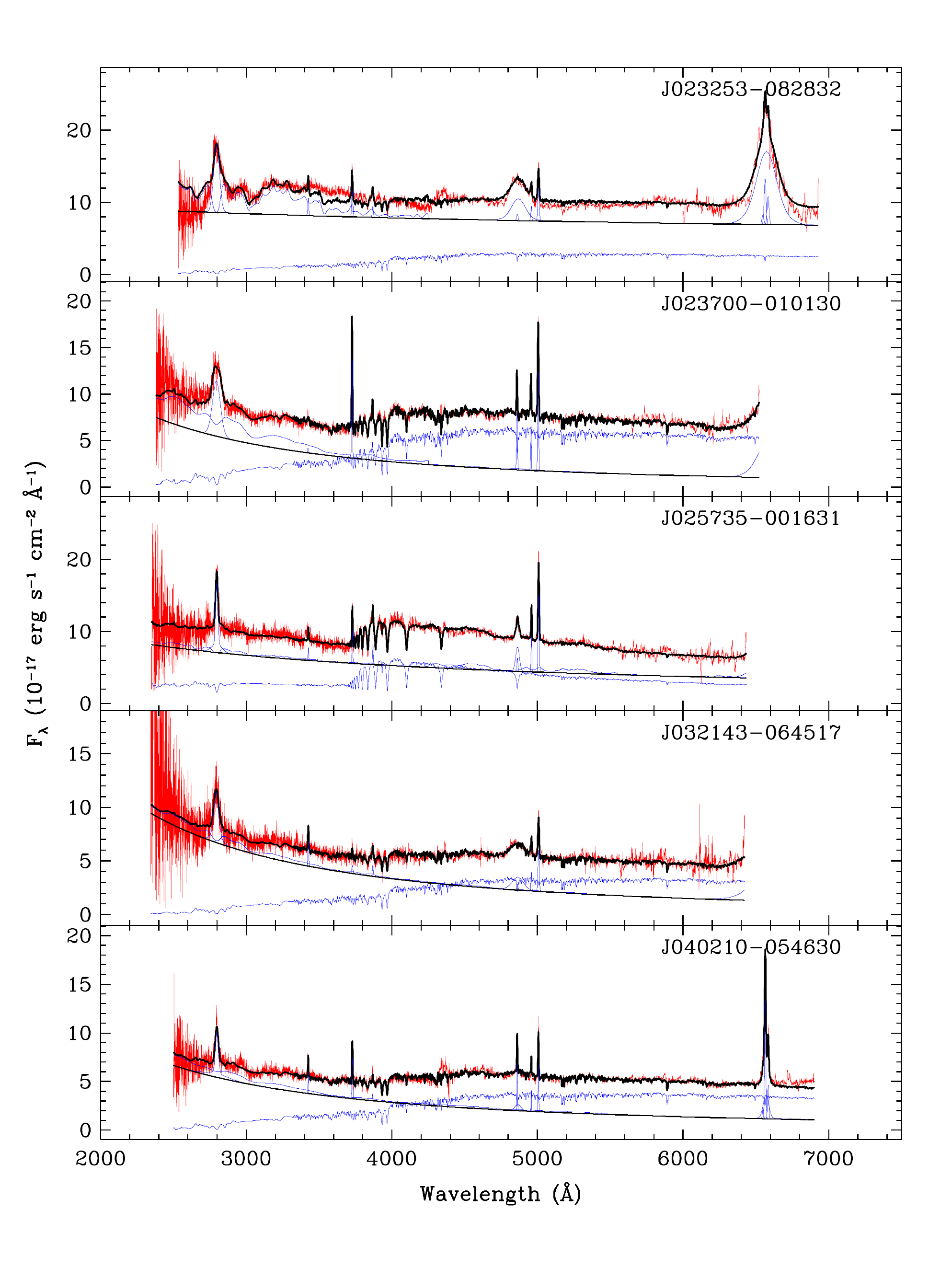}}
      \end{figure}
\begin{figure}
   \ContinuedFloat
   \centering
      \subfloat[][Fig. 8$c$]{\includegraphics[width=6in]{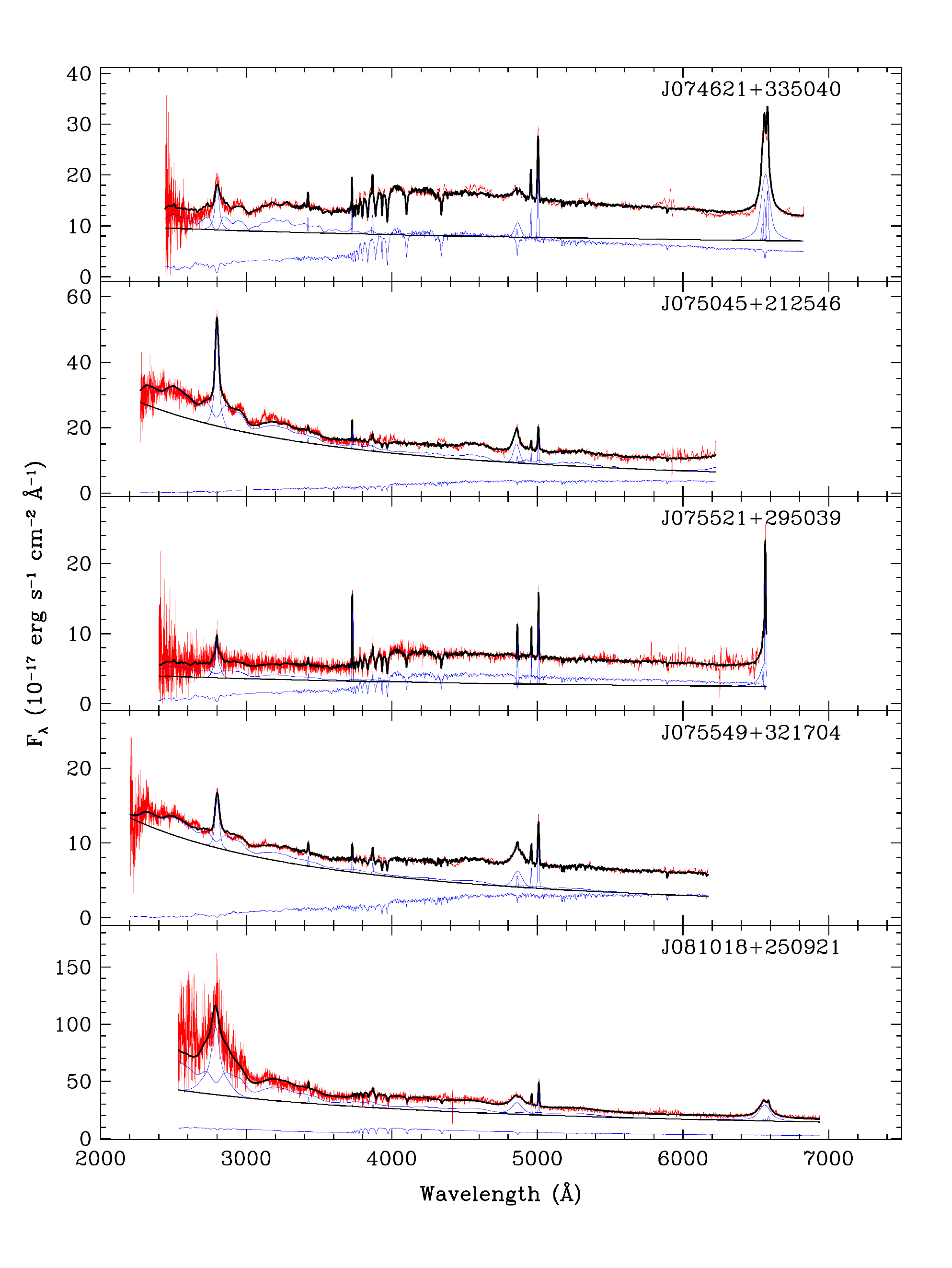}}
      \end{figure}
\begin{figure}
   \ContinuedFloat
   \centering
      \subfloat[][Fig. 8$d$]{\includegraphics[width=6in]{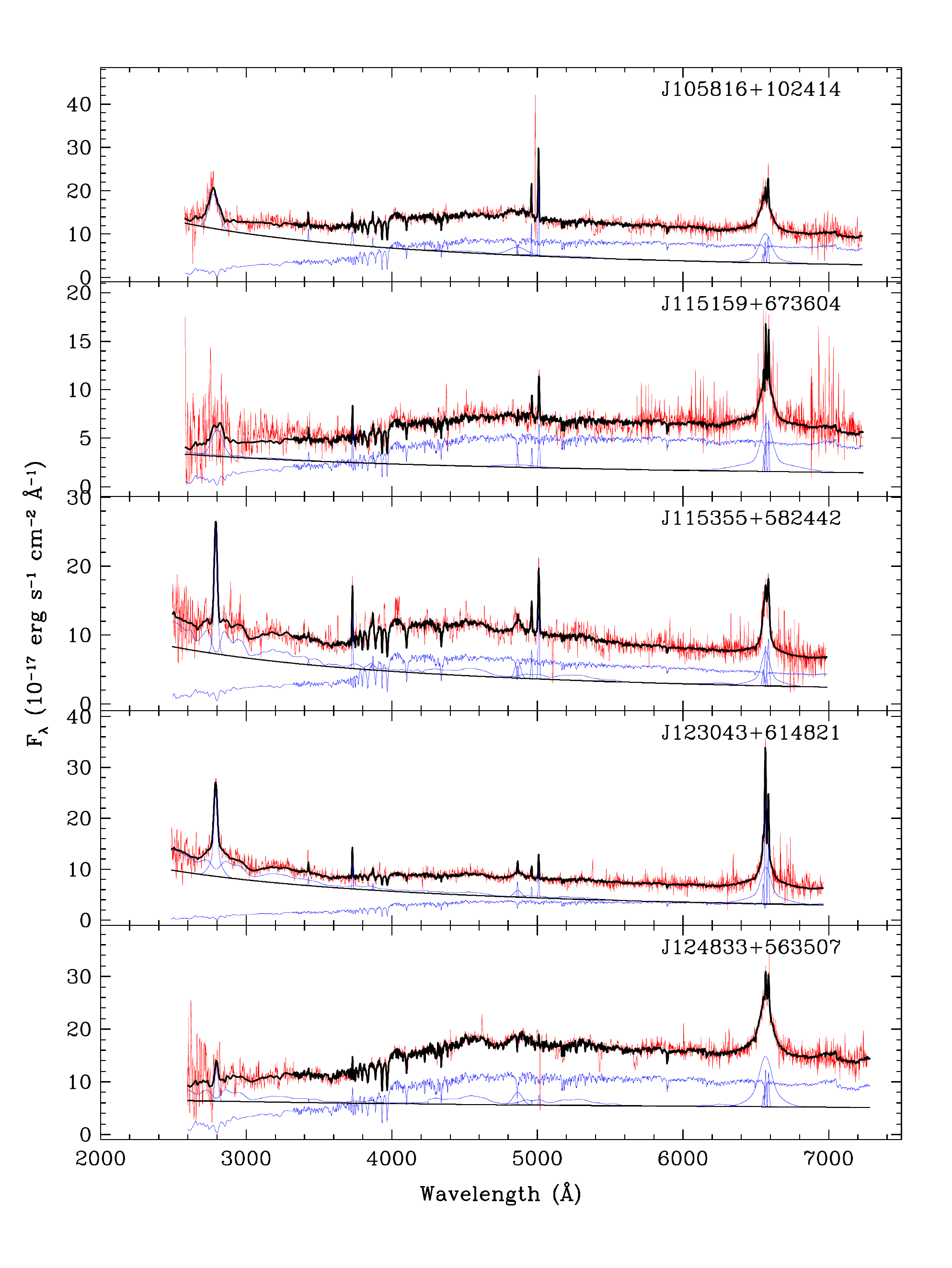}}
      \end{figure}
\begin{figure}
   \ContinuedFloat
   \centering
      \subfloat[][Fig. 8$e$]{\includegraphics[width=6in]{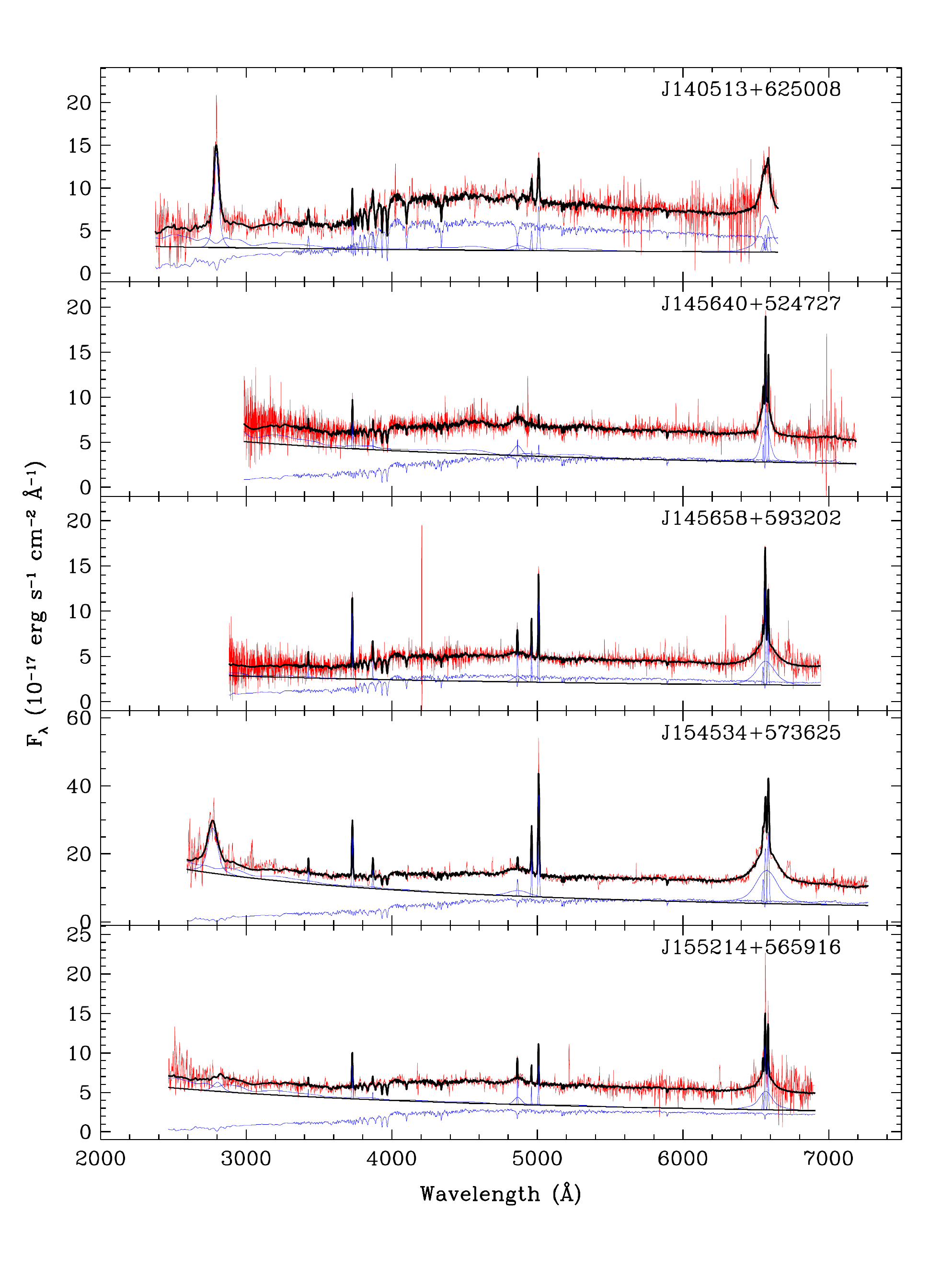}}
      \end{figure}
\begin{figure}
   \ContinuedFloat
   \centering
      \subfloat[][Fig. 8$f$]{\includegraphics[width=6in]{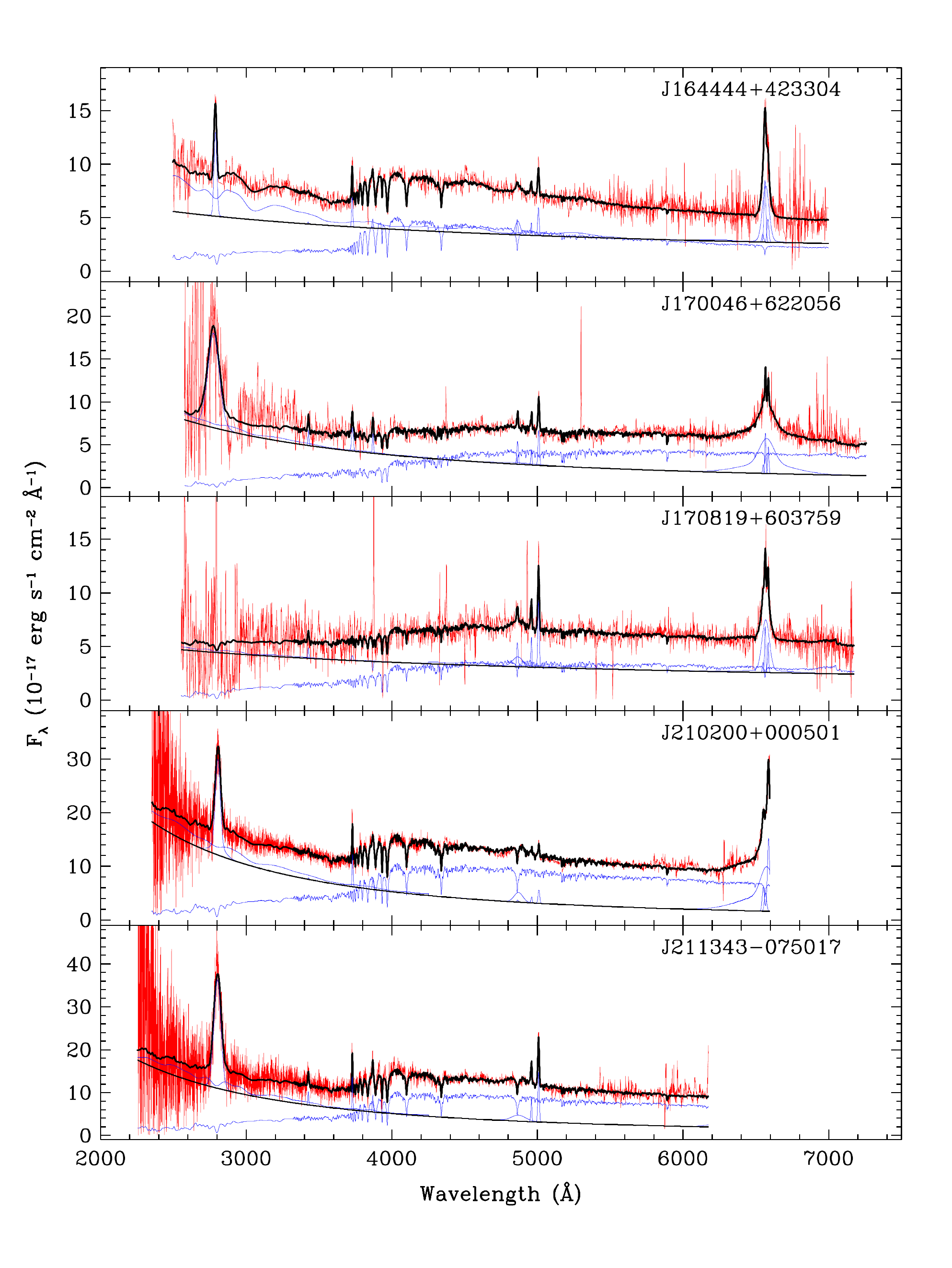}}
      \end{figure}
\begin{figure}
   \ContinuedFloat
   \centering
      \subfloat[][Fig. 8$g$]{\includegraphics[width=6in]{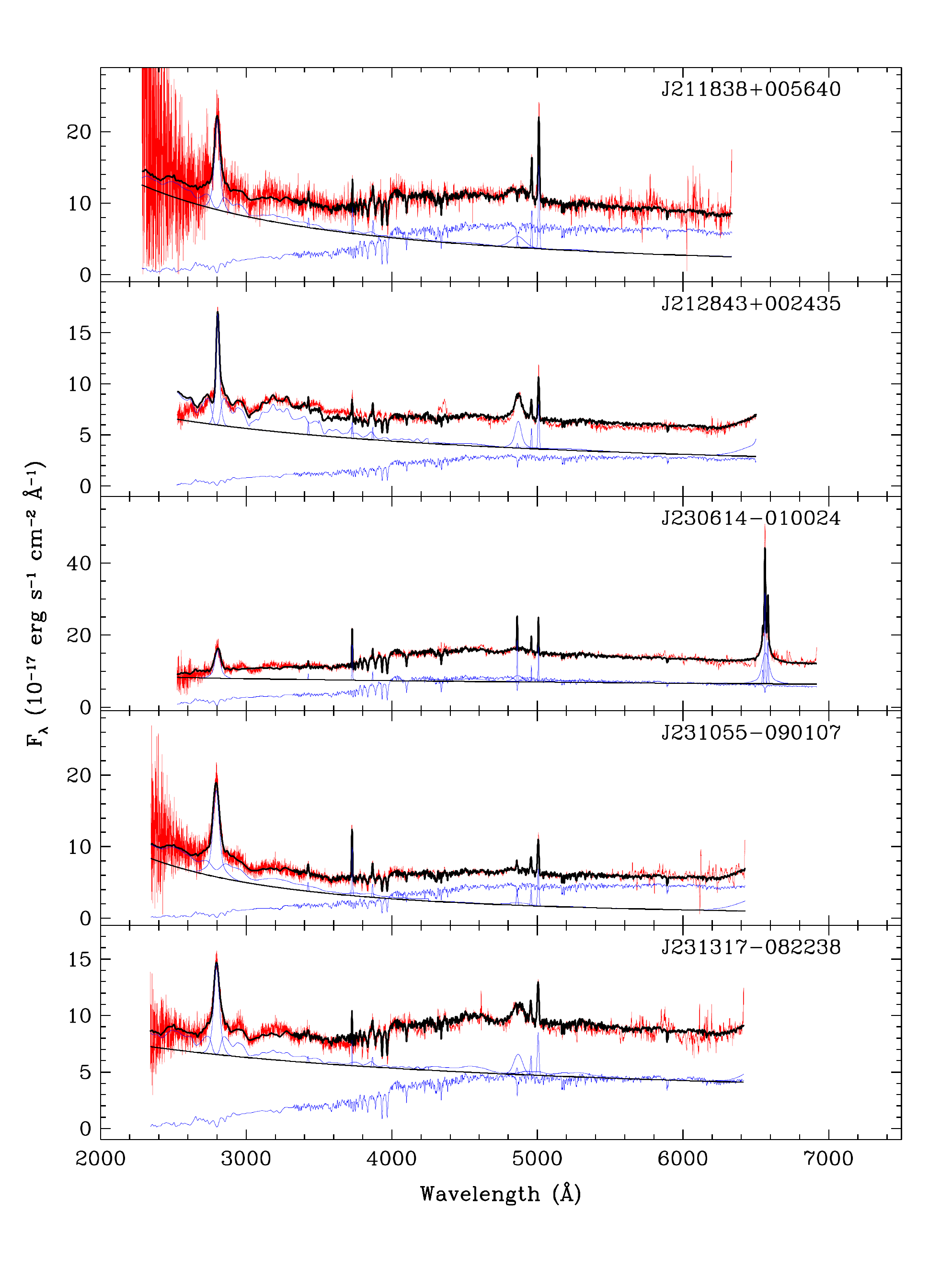}}
      \end{figure}
\begin{figure}
   \ContinuedFloat
   \centering
      \subfloat[][Fig. 8$h$]{\includegraphics[width=6in]{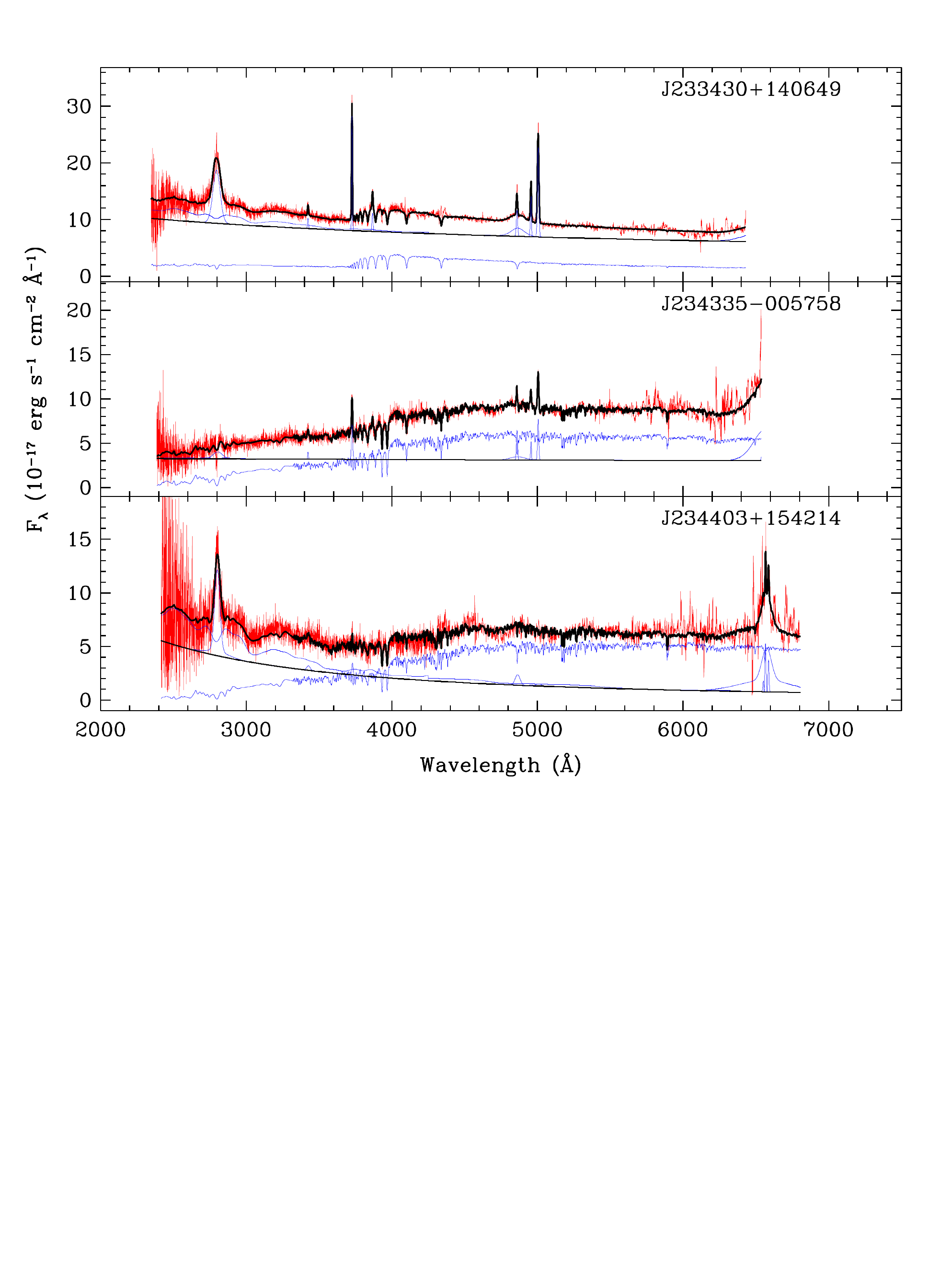}}
\end{figure}

\clearpage

\end{document}